\documentclass[%
reprint,
nofootinbib,
 amsmath,amssymb,
 aps,
 prd,
floatfix]{revtex4-2}

\usepackage{graphicx}
\usepackage{adjustbox}
\usepackage{bm}
\usepackage{aas_macros}
\usepackage{multirow}
\usepackage{tabularray}
\usepackage{booktabs}
\usepackage{xcolor}
\usepackage{cprotect}

\newcommand{\thetagauss}{\theta_{\rm FWHM}}

\begin{document}

\preprint{APS/123-QED}

\title{Constraining cosmology with thermal Sunyaev-Zel'dovich maps: \\ Minkowski functionals, peaks, minima, and moments}
\author{Alina Sabyr$^{1}$}
\email{as6131@columbia.edu}
\author{J.~Colin Hill$^{2}$}%
\author{Zolt\'an Haiman$^{1,2}$}
\affiliation{$^{1}$Department of Astronomy, Columbia University, New York, NY 10027, USA}
\affiliation{$^{2}$Department of Physics, Columbia University, New York, NY 10027, USA}

\date{\today}
\begin{abstract}
The thermal Sunyaev-Zel'dovich effect (tSZ) is a sensitive probe of cosmology, as it traces the abundance of galaxy clusters and groups in the late-time Universe. Upcoming cosmic microwave background experiments such as the Simons Observatory (SO) and CMB-S4 will provide low-noise and high-resolution component-separated tSZ maps covering a large sky fraction. The tSZ signal is highly non-Gaussian;  therefore, higher-order statistics are needed to optimally extract information from these maps. In this work, we study the cosmological constraining power of several tSZ statistics --- Minkowski functionals (MFs), peaks, minima, and moments --- that have yielded promising results in capturing non-Gaussian information from other cosmological data. Using a large suite of halo-model-based tSZ simulations with varying $\Omega_{c}$ and $\sigma_{8}$ (154 cosmologies and over $800, 000$ maps, each $10.5\times10.5$ deg$^{2}$), we show that by combining these observables, we can achieve $\approx 29\times$ tighter constraints compared to using the tSZ power spectrum alone in an idealized noiseless case, with the MFs dominating the constraints. We show that much of the MF constraining power arises from halos below the detection threshold of cluster surveys, suggesting promising synergies with cluster-count analyses. Finally, we demonstrate that these statistics have the potential to deliver tight constraints even in the presence of noise. For example, using post-component-separation tSZ noise expected for SO, we obtain $\approx1.6\times$ and $\approx1.8\times$ tighter constraints than the power spectrum with MFs and all statistics combined, respectively. We show that the constraints from MFs approach the noiseless case for white-noise levels $\lesssim 1 \,\, \mu$K-arcmin.
\end{abstract}
\maketitle
\section{Introduction}

Cosmic microwave background (CMB) primary anisotropies have allowed us to place extraordinarily tight constraints on the parameters characterizing the $\Lambda$CDM model  (e.g., \emph{WMAP} \cite{Spergel2003}, \emph{Planck} \cite{Planck2020}). Current and next-generation high-resolution multi-frequency CMB anisotropy experiments such as the Atacama Cosmology Telescope (ACT)\footnote{\url{https://act.princeton.edu/}}, South Pole Telescope (SPT)\footnote{\url{https://pole.uchicago.edu/public/Home.html}}, Simons Observatory (SO)\footnote{\url{https://simonsobservatory.org/}}, and CMB-S4\footnote{\url{https://cmb-s4.org/}}, are now also able to measure the `secondary anisotropies' --- those arising from processes after recombination --- to high precision.

One of these secondary effects, which is amongst the leading contributions to the CMB temperature fluctuations on arcminute scales, is the thermal Sunyaev-Zel'dovich (tSZ) effect --- a spectral distortion in the CMB produced via the inverse-Compton scattering of CMB photons off free, energetic electrons, primarily located in galaxy groups and clusters \cite{Zeldovich1969,Sunyaev1970}. Under the assumption of self-similarity, the observed tSZ effect scales with dark matter halo mass as $\propto M^{5/3}$~\cite{Kaiser1986}.

The unique spectral signature of the tSZ effect and its sensitivity to the abundance of massive halos --- independent of redshift --- make it a powerful cosmological probe. It can be used to identify clusters in blind surveys, providing nearly mass-limited samples, which can then be used to constrain cosmological parameters via the halo mass function (e.g., \cite{Haiman2001clusters, Allen2011,Hasselfield2013clusters, Planck2014clusters, Zubeldia2019clusters, Bocquet2024clusters}). The signal can also be studied statistically across the sky through CMB temperature maps at a given frequency or component-separated tSZ maps via multi-frequency CMB observations (e.g., \cite{KomatsuSeljak2002,Bhattacharya2012, Wilson2012, HillPajer2013, HillSherwin2013, planck_power_spectrum2014, Hill2014PDF, HorowitzSeljak2017, HurierLacasa2017,Salvati2018, Bolliet2018}). In this paper, we will focus on the latter, statistical approach.

So far, the tSZ power spectrum (e.g., \cite{KomatsuSeljak2002,HillPajer2013,Planck2014szps, HorowitzSeljak2017,HurierLacasa2017,Salvati2018,Bolliet2018}), bispectrum (e.g., \cite{Bhattacharya2012,Planck2014szps}), one-point probability distribution (PDF) (e.g., \cite{Hill2014PDF,Planck2014szps,Thiele2019}), moments (up to fourth order, e.g., \cite{Wilson2012, HillSherwin2013}) and skew-spectra (e.g., \cite{Munshi2012tSZ, Munshi2013skew}) have been studied. With respect to the cosmological parameters, the tSZ signal has been shown to have high sensitivity to the amplitude of linear matter density fluctuations at $z=0$, $\sigma_{8}$, through the steep dependence of the strength of the signal on halo mass, and therefore strong response to the halos in the exponential tail of the mass function (e.g., \cite{KomatsuSeljak2002, Bhattacharya2012, HillSherwin2013}). The tSZ signal is also sensitive to the dark energy equation of state parameter, $w$, massive neutrinos, $M_{\nu}$, and primordial non-Gaussianity through their effects on the growth of large-scale structure (e.g., \cite{HillPajer2013, Bolliet2018, Bolliet2020}).

\begin{figure*}
    \centering
    \includegraphics[width=0.7\textwidth]{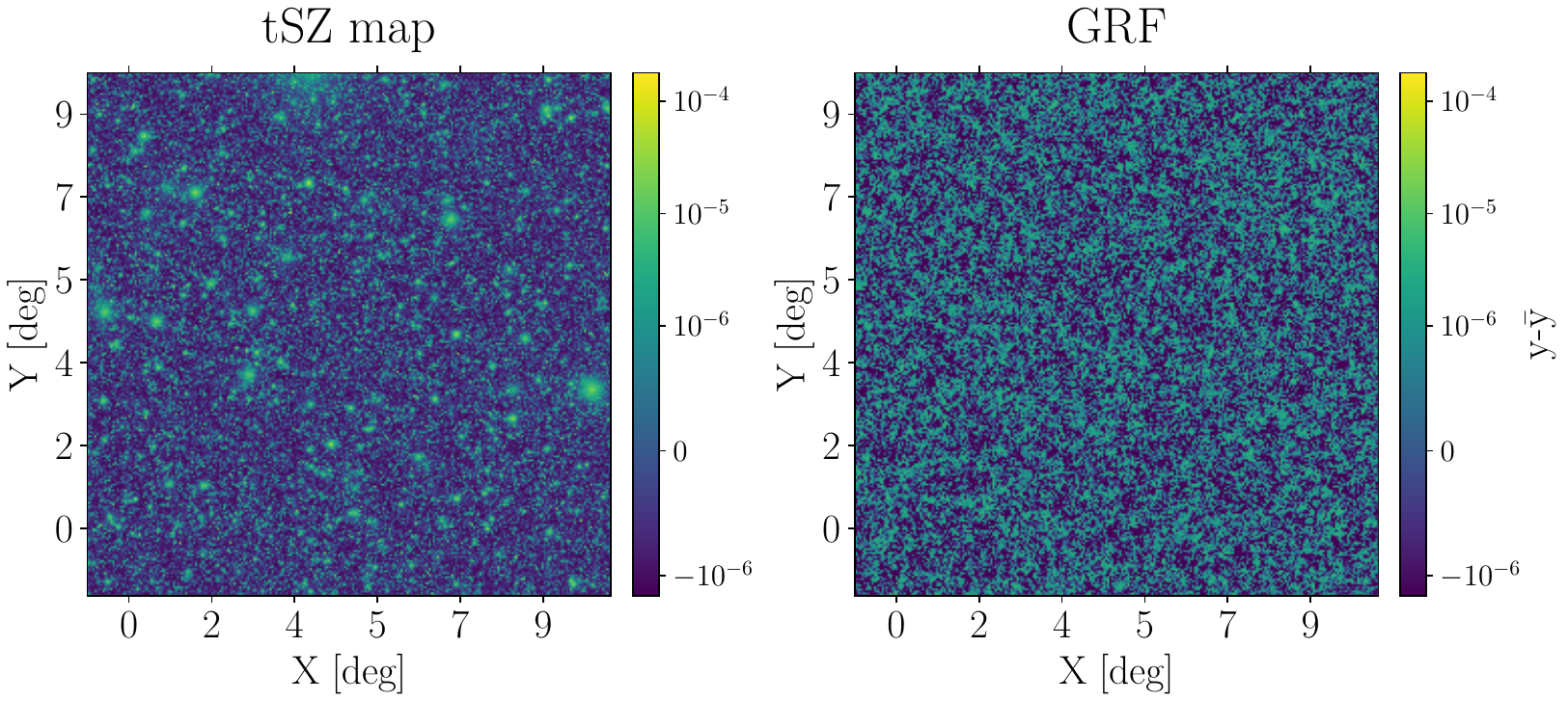}
    \caption{Comparison of a simulated halo-model-based tSZ map (at 0.1 arcmin resolution) to a Gaussian random field generated from its power spectrum. Notice the ``hot spots'' on the left, missing in the Gaussian case.}
    \label{fig:gauss_map1}
\end{figure*}

The tSZ field is highly non-Gaussian since it is sourced by massive structures in the late-time Universe (see Fig.~\ref{fig:gauss_map1}). This motivates the use of summary statistics beyond the power spectrum. Indeed, it has been shown that higher-point tSZ statistics scale more steeply with cosmological parameters \cite{HillSherwin2013, Bhattacharya2012} and that the one-point PDF provides tighter constraints than the skewness, for example~\cite{Hill2014PDF,Wilson2012}.

The primary goal of this work is to perform a systematic study of several summary statistics as a step towards: (1) finding an optimal tSZ statistic and (2) characterizing the full information content of the tSZ field. In this paper, we focus on two cosmological parameters: the cold dark matter density, $\Omega_{c}$, and $\sigma_{8}$. In particular, we compare the cosmological constraints achieved using the power spectrum to those obtained from four sets of observables: Minkowski functionals, peaks, minima, and moments.  To our knowledge, this is the first time that most of these statistics have been applied to  tSZ maps. These exact descriptors have been shown to effectively extract non-Gaussian information from other cosmological fields such as, for example, weak gravitational lensing convergence maps (e.g., \cite{Kratochvil2010peaks, Kratochvil2012MF, Petri2013, Petri2015CFHT_MF, Liu2015peaks, Marques2018, Coulton2020, Zurcher2021, Gatti2022moments, Liu2022peaks, Lanzieri2023, Marques2023peaksminima, Novaes2024, Armijo2024}).\footnote{Some of these summary statistics have also been applied to galaxy clustering (e.g., \cite{Hikage2003galaxy}) and have been used to place limits on primordial non-Gaussianity from primary anisotropies (e.g., \cite{Komatsu2009MFs, Hikage2008cmb}).} We note that the tSZ effect depends on the details of the gas physics and, in practice, parameters associated with the intracluster medium (ICM) physics would need to be simultaneously fit along with cosmological parameters. However, we focus on cosmology here and leave the exploration of the uncertainty and sensitivity related to astrophysical parameters to follow-up work.

The steep scaling of the tSZ effect with halo mass poses both a computational challenge and an advantage. In order to properly capture the sample variance and the mean for each of these summary statistics, a large number of realizations and simulation size are needed. On the other hand, this also means that the signal is dominated by the 1-halo or the Poisson term rather than the clustering of halos \cite{KomatsuKitayama1999}. For example, it has been shown for the power spectrum that the clustering effects are negligible for $\ell>300$ \cite{HillPajer2013} and are also small for the one-point PDF \cite{Thiele2019}. The effect is expected to be even less important for higher-point statistics, which scale more steeply with mass. Therefore, it is not as crucial to simulate realistic clustering of halos unlike for other cosmological fields such as, for example, weak gravitational lensing \cite{Thiele2020,Sabyr2022}, and simplified simulation approaches can be sufficiently accurate. 

Since it is time-prohibitive to perform an exhaustive study using a large suite of hydrodynamical simulations and the Poisson term dominates the tSZ signal, we make use of ``simplified simulations'' in this work. Namely, we simulate the tSZ maps by Poisson sampling halos from a given mass function, assigning a $y$-profile, and placing halos randomly in a given map \cite{Hill2014PDF,Thiele2019}. Using this fast approach, we are able to generate a large suite of simulations with sufficient size and number of random realizations to reach convergence across all the observables and to properly estimate the covariance matrices in this work. Using this suite of simulations, we study the constraining power of each of the five summary statistics both in noiseless and noisy cases. For the latter, we use post-component-separation tSZ noise power spectra for upcoming CMB experiments.

Our main findings are summarized in Fig.~\ref{fig:noiseless_all} (noiseless, Table \ref{tab:noiseless_constraints}) and Fig.~\ref{fig:noisy_main} (noisy, Table \ref{tab:noisy_constraints}). In the noiseless case, we find that combining all five summary statistics gives $\approx 29\times$ tighter constraints than using the power spectrum alone, with MFs significantly outperforming all other descriptors. The power spectrum + MFs and power spectrum + peaks yield $23\times$ and $3.4\times$ tighter constraints than the power spectrum alone, respectively. In the noisy case, we can achieve $\approx1.8\times$ tighter constraints with all the descriptors, as compared to those from the power spectrum alone. MFs still perform better than any other summary statistic: the power spectrum + MFs achieve $\approx1.7\times$ improvement over the power spectrum alone, respectively. 

The remainder of this paper is organized as follows. We describe the details of the simulation suite in \S\ref{sec:sims}, provide an overview of the statistical descriptors in \S\ref{sec:descriptors}, and outline the inference methods in \S\ref{sec:inference}. We present the results on smoothed noiseless maps in \S\ref{sec:noiseless} and the results on noisy maps in \S\ref{sec:noisy}. We summarize our findings and conclude in \S\ref{sec:discuss_conclusion}.

We assume a flat $\Lambda$CDM cosmology throughout this paper with parameters set to be consistent with \emph{Planck} \cite{Planck2020}, except that we assume no massive neutrinos: $h=0.674$, where $h \equiv H_{0}/( 100$ kms$^{-1}$Mpc$^{-1})$ and $H_{0}$ is the Hubble constant; baryon density $\Omega_{b}=0.0493$; spectral index $n_{\rm s}=0.965$, and optical depth to reionization $\tau_{\rm reio}=0.054$. We set our fiducial cosmology to have $\Omega_{c}=0.264$, and  $\sigma_{8}=0.811$.

\section{Simulations}\label{sec:sims}

In the non-relativistic limit, the CMB temperature fluctuations generated due to inverse-Compton scattering at a given sky position $\hat{\mathbf{n}}$ can be written as~\cite{Zeldovich1969,Sunyaev1970}

\begin{equation}\label{eq:tsz_gnu}
    \frac{\Delta T(\nu,\hat{\mathbf{n}})}{T_{\rm CMB}}=g(\nu)y(\hat{\mathbf{n}}), \quad \mathrm{where}\quad g(\nu)=x\coth\left(\frac{x}{2}\right)-4.
\end{equation}
\noindent Here, $\nu$ is the photon frequency, $g(\nu)$ is the tSZ spectral function, $x=(h\nu)/(k_{\rm B}T_{\rm CMB}$), $h$ is Planck's constant, $k_{\rm B}$ is Boltzmann's constant, and $T_{\rm CMB}$ is the CMB temperature today. The strength of the signal is parameterized by the Compton-$y$ parameter, which is proportional to the  integral of the electron pressure along the line of sight (LOS):
\begin{equation}
\begin{split}
    y(\hat{\mathbf{n}})=\frac{\sigma_{T}}{m_{e}c^{2}}\int n_{\rm e}(\hat{\mathbf{n}}, l)k_{B}T_{e}(\hat{\mathbf{n}},l)dl\\
    =\frac{\sigma_{T}}{m_{e}c^{2}}\int P_{e}(\hat{\mathbf{n}},l)dl
\end{split}
\end{equation}
where $n_{\rm e}$, $T_{\rm e}$, $P_{\rm e}$, and $m_{\rm e}$ are the electron number density, temperature, pressure, and mass, respectively; $\sigma_{T}$ is the Thomson cross-section, and $c$ is the speed of light. The simulated maps that we use in this work are directly in units of $y$ and we therefore denote various quantities in this work in terms of $y$. 

\subsection{tSZ maps}
To simulate the tSZ signal, we use the publicly available code \verb|hmpdf| \cite{Thiele2020,Thiele2019}\footnote{\url{https://github.com/leanderthiele/hmpdf}}. The maps are generated by Poisson sampling halos from the Tinker et al.~(2010)  \cite{Tinker2010} halo mass function in discrete bins of mass and redshift. The halo mass function is computed using the linear matter power spectrum from \verb|CLASS| \cite{CLASS2011}. Each halo is assigned a Compton-$y$ profile using the pressure profile fitting function from Battaglia et al.~(2012) \cite{Battaglia2012}. Both halo mass and the pressure profile fitting functions are defined for a spherical overdensity mass, $M_{\Delta}$, corresponding to an enclosed mass within $r_{\Delta}$ with interior density equal to ${\Delta}$ times the critical (c) or mean matter density (m) at that redshift. To convert between $M_{200\rm{m}}$ and $M_{200{\rm c}}$, the concentration-mass relation from Duffy et al.~(2008) \cite{Duffy2008} is used.  Unless otherwise noted, we set the following bounds for the mass function: $10^{11} M_{\odot}\leq M \leq 10^{16} M_{\odot}$, $0.005 \leq z \leq 6$. For the $y$-profiles, we apply a radial cutoff at $r_{\rm out}=2 r_{\rm vir}$, where $r_{\rm vir}$ is the virial radius. We use 150 bins in both redshift and mass (i.e., $N_{z} = N_{M}=150$), with bins spaced according to Gauss-Legendre quadrature within $(z_{\rm min}$, $z_{\rm max})$ and $(\ln M_{\rm min}$, $\ln M_{\rm max})$ to populate our maps, based on the convergence of the mean value of each of the summary statistics applied in this work (see Appendix~\ref{sec:convergence} for a description of the simulation convergence tests).

For computational speed, we simulate random realizations of maps that cover 10$\%$ of the sky at 0.1 arcmin resolution and cut each large map into 36 smaller ones. To remove any possible correlations at the edges, we trim 50 pixels on each side of the small maps. The final size of each map is $10.535\times10.535$ deg$^{2}$ ($6321\times6321$ pixels$^{2}$). We apodize each small map with a cosine function, using 5 pixels for the width of the apodization at each edge, and then smooth the map using a Gaussian kernel with $\thetagauss=1.4$ arcmin. An example simulated, smoothed map is shown in panel~(1) of Fig.~\ref{fig:map_noise_filter}.

We choose the pixel resolution to be 0.1 arcmin based on a comparison of the high-$\ell$ tSZ power spectra, $C_{\ell}^{yy}$, between maps generated at several different resolutions and smoothed with the same Gaussian kernel. To validate our simulations, we compare the power spectrum estimated from the maps to that computed from the halo-model code \verb|class_sz| \cite{class_sz_mm_paper,Bolliet2018,Bolliet2020,Bolliet2023,KomatsuSeljak2002}\footnote{\url{https://github.com/CLASS-SZ/class_sz}} at the fiducial cosmology (see Appendix~\ref{sec:convergence}).

Since in this work we only use the power spectra estimated from the simulated maps and, therefore, are only sensitive to the relative difference between different sets of simulations, we do not divide out the pixel window function or account for the apodization mask in the forecasts. Our map pipeline, therefore, only includes apodization and smoothing operations. We also note that we always subtract the mean value from the final maps over which we measure each of the statistics; thus, the values in the maps are both negative and positive.

\subsection{Simulation suite}
\begin{figure}
    \centering
    \includegraphics[width=\linewidth]{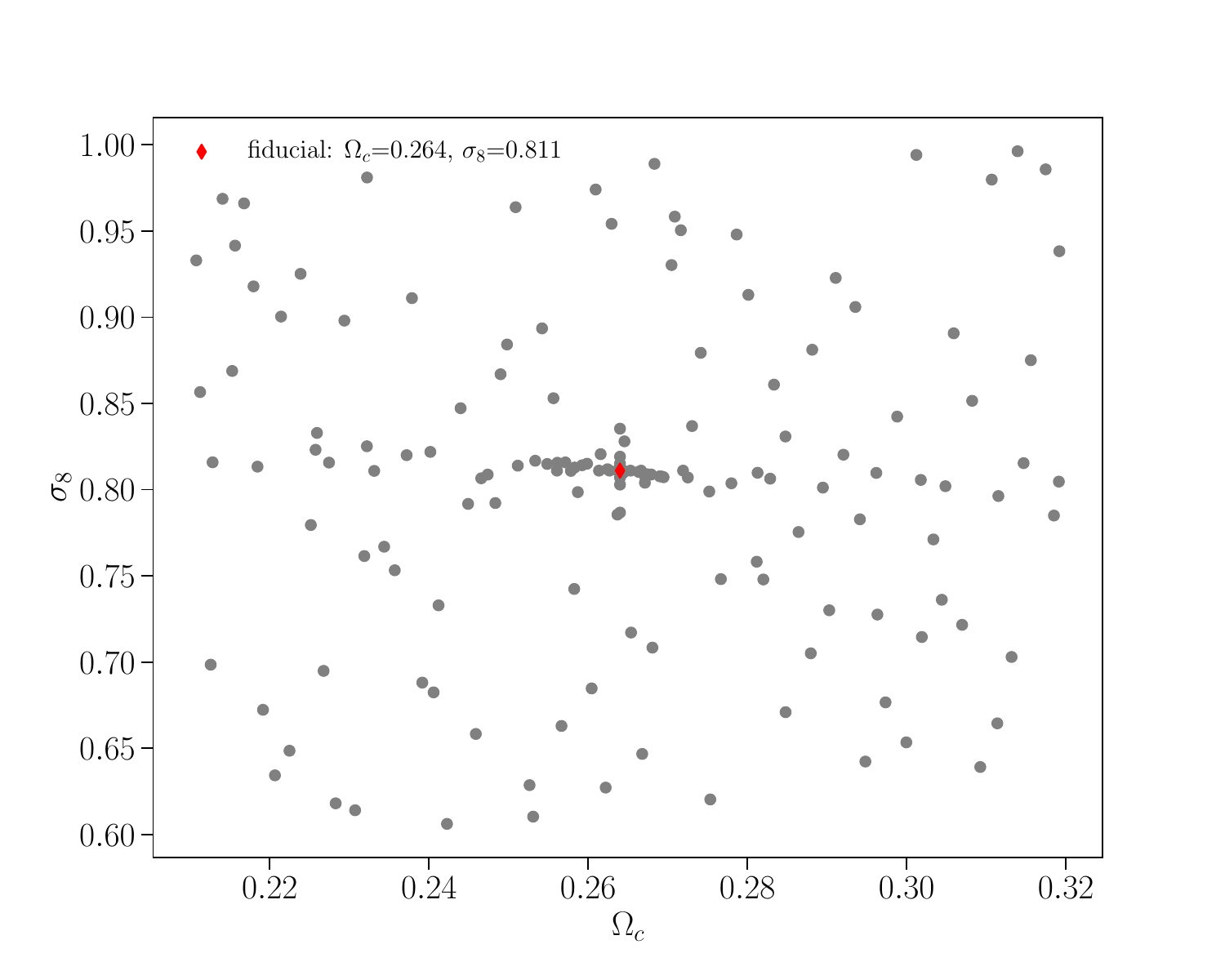}
    \caption{The simulation suite used in this work consists of 154 cosmologies with varying $\Omega_{c}$ and $\sigma_{8}$. The points are sampled more densely near the fiducial cosmology. We have $N_{\rm fid}=34,560$ random realizations at the fiducial cosmology (red diamond marker) and $N_{\rm sims}=5,112$ at every other cosmology (grey points).}
    \label{fig:suite}
\end{figure}

Our simulation suite consists of 154 different cosmologies, as shown in Fig.~\ref{fig:suite}, with varying parameters $\Omega_{c}$ and $\sigma_{8}$, with $N_{\rm fid}=34,560$ random realizations at the fiducial cosmology and $N_{\rm sims}=5,112$ at all others. We set our fiducial cosmology to be at $\Omega_{c}=0.264, \sigma_{8}=0.811$ and fix the rest of the cosmological parameters to results from \emph{Planck} \cite{Planck2020}, but with no massive neutrinos: $h=0.674$, $\Omega_{b}=0.0493$, $n_{\rm s}=0.965$, $\tau_{\rm reio}=0.054$. 

We choose the number of simulations at the fiducial cosmology based on an estimate of how many realizations will be needed to accurately compute the covariance for a summary statistic with dimension $d\sim$ few hundred (i.e., $N_{\rm fid} \gtrsim d^{2}/2$). In \S.\ref{sec:noiseless_convergence}, we show that the constraints are stable with $N_{\rm fid}<34,560$. We pick $N_{\rm sims}=5,112$ for other cosmologies based on the stability of the mean value of each summary statistic considered in this work. As described in Appendix~\ref{sec:convergence}, the mean converges across the mass function discretization settings in \verb|hmpdf| at $N_{\rm sims}<5,112$.

We create the suite by first running 102 cosmologies within the broad ranges $\Omega_{c}\in\{0.21,0.32\}$ and  $\sigma_{8}\in\{0.6,1\}$ (approximately 20$\%$ or $30\sigma$ from the \emph{Planck} central values, respectively) with 99 of those sampled according to the Latin hypercube (LH). After computing the initial constraints from all the statistics, we add extra cosmologies in the region near the maximum likelihood to ensure that we have a sufficient number of cosmologies to interpolate each of the summary statistics (see \S~\ref{sec:inference}) and obtain reliable contours. In particular, we add 20 cosmologies sampled along a line fit to the $\Omega_c-\sigma_8$ degeneracy found for the power spectrum and 20 additional points around the constraints from all the statistics combined. The latter we do by following a similar approach as in Ref.~\cite{ZM2016}. We first compute the approximate covariance using the parameter values that lie within the $1\sigma$ constraints and the confidence ellipse parameters $a,b,\gamma$ following Ref.~\cite{Coe2009}:
\begin{equation}\label{eq:ellipse_params}
\begin{split}
    a^{2}=\frac{\sigma_{x}^{2}+\sigma_{y}^2}{2}+\sqrt{\frac{(\sigma_{x}^{2}-\sigma_{y}^2)^{2}}{4}+\sigma_{xy}^{2}}\\
    b^{2}=\frac{\sigma_{x}^{2}+\sigma_{y}^2}{2}-\sqrt{\frac{(\sigma_{x}^{2}-\sigma_{y}^2)^{2}}{4}+\sigma_{xy}^{2}}\\
    \tan(2\gamma)=\frac{2\sigma_{xy}}{\sigma_{x}^{2}-\sigma_{y}^{2}}
\end{split}
\end{equation}
We then sample values for $r \in \{0,1\}$, $\phi\in\{0,2\pi\}$ and convert to $\Omega_{c}, \sigma_{8}$ via
\begin{equation}
\begin{split}
x = \alpha a\cos(\phi)r^{n}\\
y = \alpha b\sin(\phi)r^{n}\\
\Omega_{\rm c}=x\cos(\gamma)-y\sin(\gamma)+\Omega_{\rm c, fid}\\
\sigma_{8}=x\sin(\gamma)+y\cos(\gamma)+\sigma_{\rm 8, fid}\\
\end{split}
\end{equation}
where we set $n=1$, $\alpha=3.5$ ($\alpha=3.44$ corresponds to 3$\sigma$). Choosing $n > 1/2$ means that the density of samples increases towards the center. In Appendix.~\ref{app:interp_bin}, we show that with these cosmologies we are able to construct an accurate interpolator for all the summary statistics and achieve stable forecasts.

Finally, we add 12 more cosmologies that can be used for Fisher forecasts, where one of the two parameters is perturbed from its fiducial value, while the other is fixed. For these we use $\delta\theta=\pm0.5,1,3 \%$, which correspond to $\delta\Omega_{c}=\pm\{0.00132,0.00264,0.00792\}$, $\delta\sigma_{8}=\pm\{0.00406,0.00811,0.02433\}$.  

\subsection{Noise}
To study the constraining power of the tSZ statistics in the presence of realistic noise, we generate Gaussian random fields (GRFs) using different noise power spectra, $N_{\ell}^{yy}$. Note that here $N_{\ell}^{yy}$ refers to the non-beam-deconvolved noise power spectra, as we always use smoothed tSZ maps to compute the forecasts.

First, we consider several experimental set-ups using the post-component-separation Compton-$y$ noise power spectra forecasted for SO (baseline and goal noise levels)\footnote{\url{https://github.com/simonsobs/so_noise_models/tree/master/LAT_comp_sep_noise/v3.1.0}} \cite{SO_noise2019} and CMB-S4 \cite{CMB-S4_noise2019}\footnote{See Fig.~20 of Ref.~\cite{CMB-S4_noise2019}.}, both computed assuming combination with \emph{Planck} data. These power spectra were generated for an assumed model, which includes both the instrumental noise and foreground components. Thus, these noise power spectra include the expected residuals from the other sky components during the component-separation process to obtain tSZ maps from multi-frequency CMB observations.  However, note that the noise power spectra (including the residual foreground fields) are Gaussian and are not correlated with the tSZ field, complexities that we leave to future work.  In all cases, we use the noise computed via the standard internal linear combination (ILC) method and assume $f_{\rm sky}=0.4$ (for more details, see Refs.~\cite{SO_noise2019,CMB-S4_noise2019}). 

\begin{figure}
    \centering
    \includegraphics[width=\columnwidth]{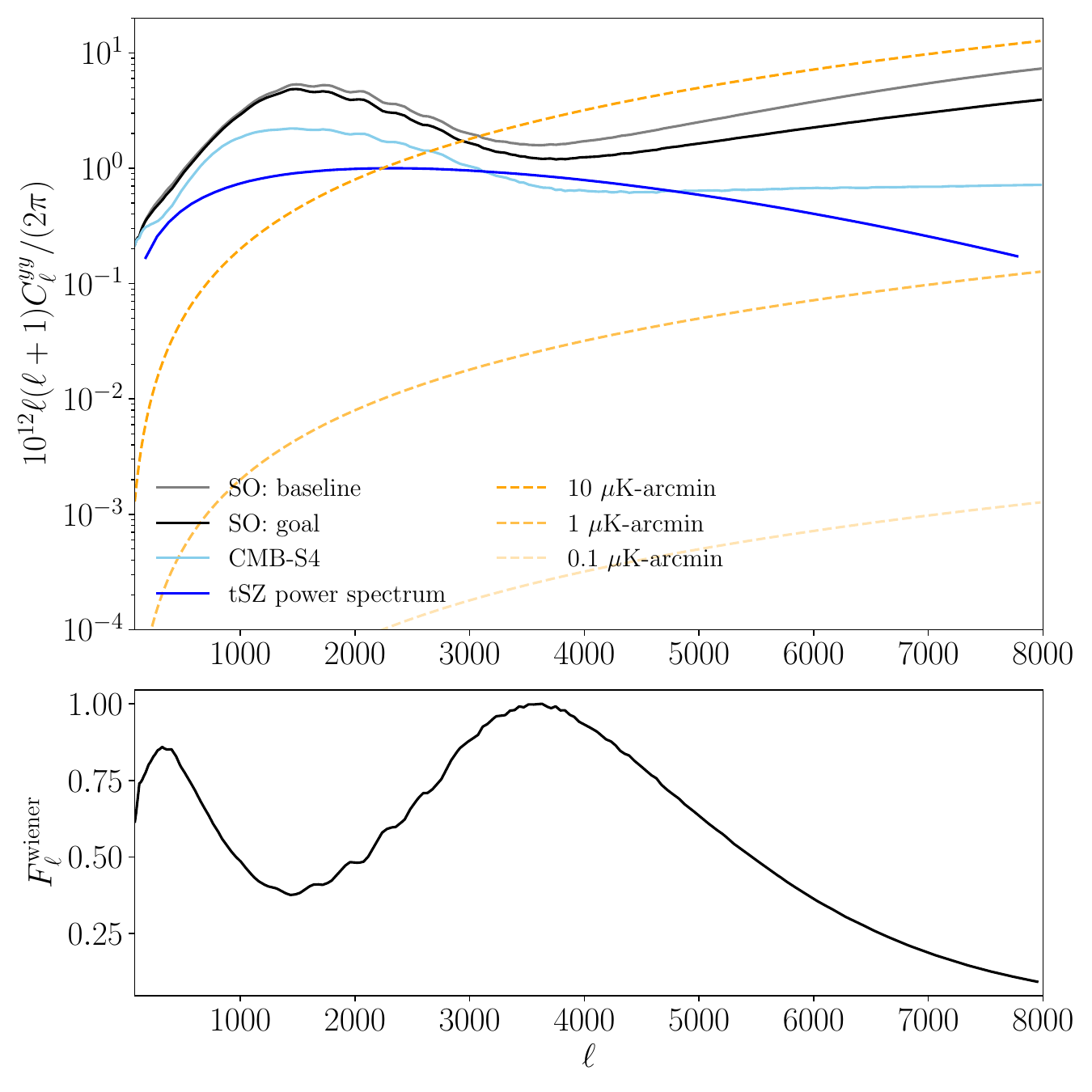}
    \caption{\textit{Top:} Post-component-separation tSZ noise for SO baseline (goal) noise levels in grey (black) and CMB-S4 (cyan) compared to the tSZ power spectrum (blue) computed at the fiducial cosmology. Combination with \emph{Planck} data is assumed in these noise power spectra and the increase in noise around $\ell\approx 1000-1500$ shows the transition between the  \emph{Planck}- and SO/CMB-S4-dominated regimes (i.e., low-$\ell$ modes are noisy for SO/CMB-S4 due to atmospheric effects). White-noise power spectra for different noise levels are plotted with dashed curves in orange.  Note that the signal curve here includes smoothing with a FWHM = 1.4 arcmin beam. \textit{Bottom:} Wiener filter computed for the SO goal noise.}
    \label{fig:noise}
\end{figure}

The largest-angular-scale noise is not well-captured by the ILC method due to the small number of harmonic modes at low $\ell$. To avoid any ill behavior in the maps due to extrapolation of the noise power spectra at low- and high-$\ell$ ranges, we filter out the scales $\ell\leq80$ and $\ell\geq7950$ (smoothly tapering with a $\tanh$ function between $\ell=80-90$ and $\ell=7940-7950$) from both the tSZ maps and the noise maps (i.e., near the boundaries of the SO noise power spectra). The noise curves and their comparison to the tSZ power spectrum at the fiducial cosmology are shown in Fig.~\ref{fig:noise}.

\begin{figure*}
    \centering
    \includegraphics[width=\textwidth]{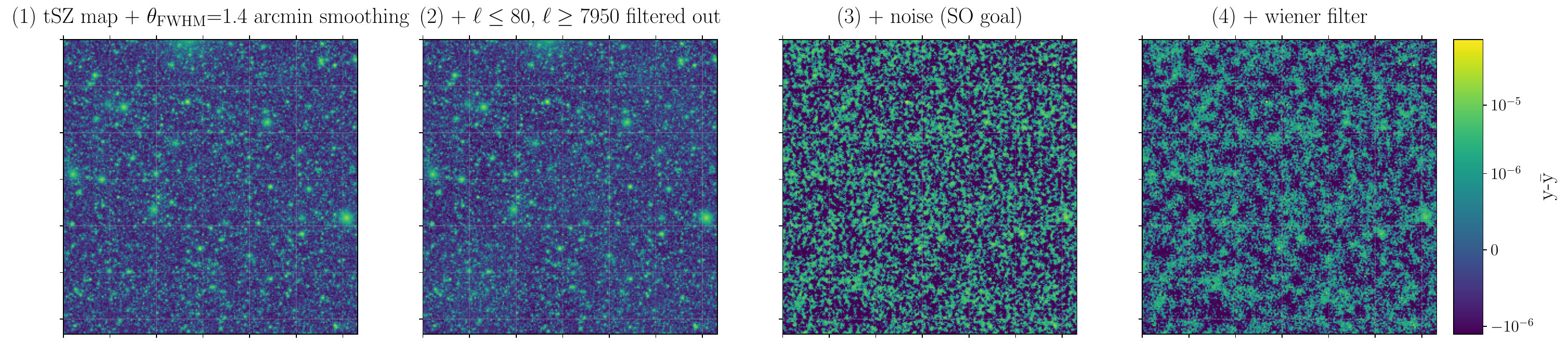}
    \caption{Simulated maps used to obtain various forecasts in this work at different steps of processing. (1) Noiseless tSZ map smoothed with a Gaussian kernel with $\theta_{\rm FWHM}=1.4$ arcmin, (2) low- and high-$\ell$ modes filtered out, (3) adding a Gaussian random field generated using the post-component-separation tSZ noise power spectra forecasted for the SO goal noise levels, and (4) effect of applying a Wiener filter to the noisy map.}
    \label{fig:map_noise_filter}
\end{figure*}

Fig.~\ref{fig:map_noise_filter} shows visually the impact of adding noise to the tSZ maps for the SO goal case. Panel (1) is the noiseless map described in the previous section smoothed with a $\thetagauss=1.4$ arcmin Gaussian kernel. Panel (2) shows the effects on the tSZ map of filtering out the low- and high-$\ell$ modes. Panel (3) is the total map, comprised of the smoothed tSZ map plus the random noise realization. While some structure from the tSZ field still remains, the noisy map is much more similar to a GRF (see, e.g., Fig.~\ref{fig:gauss_map1}).

To mitigate the effects of noise in the real-space statistics that we study, we apply a Wiener filter to the noisy tSZ maps. We compute the $\ell$-space filter using the tSZ power spectrum measured at the fiducial cosmology (averaged over $N_{\rm fid}=34,560$ realizations) from the noiseless maps and $N_{\ell}^{yy}$ using $80<\ell<7950$:
\begin{equation}
\label{eq:wiener}
    F^{\rm Wiener}_{\ell}=\frac{C_{\ell}^{yy}}{C_{\ell}^{yy}+N_{\ell}^{yy}} \,.
\end{equation}
The filter preferentially upweights the modes that contain the largest tSZ signal-to-noise ratio. Since we construct the filter using the power spectrum, it is not guaranteed that it is optimal for the real-space non-Gaussian statistics that we analyze. Nonetheless, it still suppresses the modes that are clearly dominated by noise. We plot the filter for SO goal noise levels in the bottom panel of Fig.~\ref{fig:noise} and show its effects on the simulated maps in the fourth panel of Fig.~\ref{fig:map_noise_filter}. We normalize the filter via dividing it by its maximum value.

In addition to noise power spectra from realistic experimental set-ups, we also consider different levels of white noise: $\Delta T=0.1, 1,$ and $10$ $\mu$K-arcmin. To generate GRFs, we convert from CMB temperature units, $\mu$K-arcmin, to Compton-$y$ units using the tSZ spectral function $g(\nu)$ in Eq.~\eqref{eq:tsz_gnu} and $T_{\rm CMB}=2.7255$ K. We convert the units at $\nu=150$ GHz. The white noise maps are then generated assuming a flat power spectrum with an amplitude equal to the noise level in Compton-$y$ units $\times$ radians squared. As in the case of the post-component noise power spectra, we apply the same low- and high-$\ell$ filter and a Wiener filter computed for each of the flat noise power spectra cases. We show the white noise power spectra in Fig.~\ref{fig:noise} and the combined tSZ plus white noise maps in Fig.~\ref{fig:white_noise}. Note that for white noise levels of 0.1 or 1 $\mu$K-arcmin, the maps closely resemble the noiseless case, as expected from the signal versus noise power spectra comparison shown in Fig.~\ref{fig:noise}.

\section{Summary statistics}\label{sec:descriptors}
In this paper, we employ five different summary statistics: the angular power spectrum, Minkowski functionals, peaks, minima, and moments. We use the implementation in \verb|LensTools|\cite{lenstools}\footnote{\url{https://github.com/apetri/LensTools}} and describe each of them and our fiducial choices in this section. We note that some of the statistics are measured for a set of bins. To define the bins, we generally use values defined in terms of the average rms of the tSZ maps at the fiducial cosmology. These values are $\sigma_{\rm fid}=1.5\times10^{-6}$ and $\sigma_{\rm fid}^{\rm SO, goal}=1.9\times10^{-6}$, for the fiducial noiseless and noisy cases, respectively.  These values correspond to the average rms computed over 5,184 maps at the fiducial cosmology.\footnote{The rms values are computed over 5,184 map realizations based on the number of simulations used for the initial convergence tests prior to running the suite.} We summarize the fiducial choices and the size of each of the summary statistic data vectors in Table~\ref{tab:fid_bins}.

\begin{table*}[]
    \centering
    \begin{tabular}{|l|c|c|c|c|}
    \colrule
          \multicolumn{1}{|c|}{} & \multicolumn{2}{|c|}{Noiseless}& \multicolumn{2}{|c|}{Noisy}\\ 
        \cline{1-5}
         Observable & dimension & bin edges & dimension& bin edges\\
         \colrule
         power spectrum & 77 & linearly spaced, $\ell\in\{125, 7825\}$& 77& linearly spaced, $\ell\in\{125, 7825\}$\\
         MFs & 3$\times$20 & linearly spaced, $\nu\in\{-1\sigma_{\rm fid}, 3\sigma_{\rm fid}\}$ & 3$\times$30 & linearly spaced, $\nu\in\{-3\sigma_{\rm fid}^{\rm SO, goal}, 3\sigma_{\rm fid}^{\rm SO, goal}\}$\\
         Peaks& 9 & [$-100,-0.3,-0.2,-0.1,0,0.2,$ & 9 &[$-100,-0.3,-0.2,-0.1,0,
    0.2,$\\
         & & $0.3,0.6,1,1000]\sigma_{\rm fid}$&& $0.3,0.6,1,1000]\sigma_{\rm fid}^{\rm SO, goal}$\\

         Minima & 7 & [$-100,-0.57,-0.5,-0.47,-0.42,$ &7&[$-100,-0.57,-0.5,-0.47,-0.42,$\\
    && $-0.36,-0.26,100]\sigma_{\rm fid}$&& $-0.36,-0.26,100]\sigma_{\rm fid}^{\rm SO, goal}$ \\
         Moments & 4 & -- & 4 & --\\
         \colrule
    \end{tabular}
    \caption{We list the fiducial binning choices and dimensions for each of the summary statistics both for the noiseless and noisy cases (i.e., for the results shown in Fig.~\ref{fig:noiseless_all} and Fig.~\ref{fig:noisy_main}), where $\sigma_{\rm fid}=1.5\times10^{-6}$ and $\sigma_{\rm fid}^{\rm SO, goal}=1.9\times10^{-6}$ correspond to the average rms values computed at the fiducial cosmology.}
    \label{tab:fid_bins}
\end{table*}

\subsection{Angular power spectrum}

We use the tSZ power spectrum as a baseline statistic with which to compare the constraining power of each of the summary statistics. It is defined as 
\begin{equation}
    \langle \tilde{y}_{\bm{\ell}}^{*}\tilde{y}_{\bm{\ell}'} \rangle=(2\pi)^{2}\delta_{D}({\bm{\ell}-\bm{\ell}'})C^{yy}_{\ell}
\end{equation}
where $\bm{\ell}$ is the 2D Fourier wavevector, $\tilde{y}$ indicates the Fourier transform of the field, and $\delta_{D}$ is the Dirac delta function. We average the pixel values in equal-size linearly spaced $\ell=|\bm{\ell}|$ bins between $\ell\in\{25,7925\}$ (bin width equal to $\Delta\ell=100$). We remove the lowest and highest $\ell$ bins to avoid any aliasing effects from applying an apodization mask to our maps and end up with a total of 77 bins with the first and the last bins centered at $\ell_{\rm min}=175$ and $\ell_{\rm max}=7775$. 

\subsection{Minkowski Functionals}

For a D-dimensional field, one can compute D+1 Minkowski functionals (MFs), which fully characterize the global morphological properties (i.e., invariant to rotation, translation, and additive) of that field \cite{Hadwiger1957,Minkowski1903, Schmalzing1996}\footnote{One could also study and compute the Minkowski Tensors, a tensorial generalization of MFs \cite{MF_tensor2020} (see e.g., \url{https://morphometry.org/)}.}. Thus, for a 2D tSZ map, we can measure three MFs: $V_{0}(\nu)\sim$ area, $V_{1}(\nu)\sim$ contour length, and $V_{2}(\nu)\sim$ Euler characteristic, as a function of some threshold value $\nu$. More precisely,
\begin{equation}
    V_{0}(\nu)=\frac{1}{A}\int_{\Sigma(\nu)}da,
\end{equation}
where $A$ is the total area of the map or the total number of pixels, $da$ is the area element, and $\Sigma(\nu)=\{y>\nu\}$ defines the excursion set over which the MF is measured. $V_{0}$ is, thus, the one-point cumulative distribution function (one-point CDF). Next, $V_{1}(\nu)$ is defined as 
\begin{equation}
    V_{1}(\nu)=\frac{1}{4A}\int_{\partial\Sigma(\nu)}dl,
\end{equation}
where $dl$ is the boundary length element and $\partial\Sigma(\nu)$ is the excursion set boundary. Finally, $V_{2}(\nu)$ is defined as the integral of the curvature of the boundary $\mathcal{K}$:
\begin{equation}
    V_{2}(\nu)=\frac{1}{2\pi A}\int_{\partial\Sigma(\nu)}\mathcal{K} dl
\end{equation}
In flat space, $V_{2}(\nu)$ is equivalent to the Euler characteristic \cite{Matsubara2010} and for a closed surface, it directly relates to the genus.

In practice, MFs can be efficiently computed as follows:
\begin{equation}
\begin{split}
V_{0}(\nu)=\frac{1}{A}\int_{A}\Theta(y(\mathbf{p})-\nu)didj\\
V_{1}(\nu)=\frac{1}{4A}\int_{A}\delta_{D}(y(\mathbf{p})-\nu)\sqrt{y_{i}^{2}+y_{j}^{2}}didj\\
V_{2}(\nu)=\frac{1}{2\pi A}\int_{A}\delta_{D}(y(\mathbf{p})-\nu)\\
\times \frac{2y_{i}y_{j}y_{ij}-y_{i}^{2}y_{jj}-y_{j}^{2}y_{ii}}{y_{i}^{2}+y_{j}^{2}}didj
\end{split}
\end{equation}
where $\mathbf{p}$ denotes the coordinate position $(i,j)$, $\Theta$ is the Heaviside step function, and $y_{i/j}$ is the gradient of the field with respect to coordinates $i/j$ (i.e., $y_{ii}$ corresponds to the second derivative). Thus, what is required in practice are simply the first and second derivatives of the field. In the cases where the gradients vanish (i.e., the denominator in $V_{2}$), the integrand is replaced by $(2y_{ij}-y_{jj}-y_{ii})/2$. The $\delta_{D}$-functions are approximated by dividing out $d\nu$ (see more details in Ref.~\cite{Petri2013}). We check the estimators by computing MFs on a set of GRFs for which an analytic form is known \cite{Tomita1986, Matsubara2000, Matsubara2010, Kratochvil2012MF}.

In our fiducial set-up on the noiseless maps, we measure each of the MFs using 20 linearly spaced bins between $\nu\in\{\nu_{\rm min},\nu_{\rm max}\}\in\{-1\sigma_{\rm fid}, 3\sigma_{\rm fid}\}$ (i.e., the MF data vector contains 60 elements). For the fiducial noisy case, we use $\nu\in\{\nu_{\rm min},\nu_{\rm max}\}\in\{-3\sigma_{\rm fid}^{\rm SO, goal}, 3\sigma_{\rm fid}^{\rm SO, goal}\}$ and 30 bins for each of the MFs. We choose these ranges to fully capture the shape for each of the MFs. In Appendix.~\ref{app:interp_bin}, we show that increasing the number of bins does not impact the forecasts from MFs.

\subsection{Peaks and minima}
Peaks (minima) are defined as pixels in the map that are higher (lower) than their 8 neighboring pixels. We locate the peaks and minima in the maps and bin them by their values. Our observable is then a peak (minima) count histogram. We use 9 and 7 bins for the peaks and minima, respectively, with edges defined as follows:
\begin{equation}
\begin{split}
    x^{\rm peaks}_{\rm b}=[-100,-0.3,-0.2,-0.1,0,\\
    0.2,0.3,0.6,1.,1000]\sigma_{0}\\
    x^{\rm minima}_{\rm b}=[-100,-0.57,-0.5,-0.47,-0.42,\\
    -0.36,-0.26,100]\sigma_{0}\\
\end{split} 
\end{equation}
where we use $\sigma_{0}=\sigma_{\rm fid}$ and $\sigma_{0}=\sigma_{\rm fid}^{\rm SO, goal}$ for the noiseless and noisy cases, respectively.
To choose these exact bins, we take a realization of a fiducial map and bin the peaks (minima) into 10 bins in such a way that each bin has the same number of peak (minima) counts. We adjust the bin edges to ensure the stability of the mean value of each bin with respect to the numerical settings for the maps and the number of simulations (see Appendix \ref{sec:convergence} for more details about the convergence). Although the number of peaks (minima) is expected to vary significantly across different cosmologies, we do not fine-tune the bin edges any further. As we describe in Appendix \ref{app:interp_bin}, the constraints for peaks (minima) are not very sensitive to the exact choice of the bin edges as long as at least a few bins are used. 

\subsection{Moments}
We measure four low-order moments (two quadratic, one cubic, and one quartic). This set of moments is motivated by the perturbative expansion of MFs up to second-order in variance \cite{Matsubara2000, Matsubara2010, Petri2013}. For a Gaussian random field, MFs can be expressed analytically using the variance of the field ($\sigma_{0}$) and the variance of the gradient of the field ($\sigma_{1}$). 
For non-Gaussian fields, one can write down an expansion of MFs in powers of the variance. The first-order term is described by the three skewness parameters and the second-order term by the four kurtosis parameters (see Refs.~\cite{Matsubara2000, Matsubara2010, Munshi2012}.) It has been shown that these perturbative series do not capture all the information that MFs are sensitive to for weak gravitational lensing fields unless the maps are considerably smoothed (e.g., smoothing scale $\sigma\geq 15$ arcmin) \cite{Petri2013}.
Since the tSZ field also traces late-time structure, we expect that these series do not converge for our maps using a Gaussian kernel with $\thetagauss=1.4$ arcmin. We therefore treat the MFs and moments as separate descriptors in our study as done in previous works \cite{Petri2015CFHT_MF}. The un-normalized moments that we measure are the following:
\begin{equation}
    \sigma_{0}=\sqrt{\langle y^{2} \rangle}, \quad \sigma_{1}=\sqrt{\langle |\nabla y|^{2} \rangle}
\end{equation}
\begin{equation}
    S_{1}=\langle y^{2} \nabla^{2} y\rangle,\quad K_{1}=\langle y^{3} \nabla^{2} y\rangle
\end{equation}

As in the case of measuring the MFs, we only need the gradient and the Hessian of the tSZ field to compute the moments. Skewness terms vanish in the Gaussian case, while the kurtosis terms are non-zero. In this work, we measure the full kurtosis moment and do not subtract the Gaussian (`un-connected') part. We note that the expansion of MFs up to the second-order includes two additional cubic terms and three additional quartic moments. We do not include them in our baseline results due to the large sample variance of those moments and the possible lack of convergence of their mean values with respect to the number of simulations that we use in this study. We explore the constraints from those additional moments in Appendix~\ref{app:kurtosis}.

\section{Parameter Inference}\label{sec:inference}
To assess the constraining power of each of the summary statistics described in the previous section, we follow a standard Bayesian framework to infer the posterior distributions of the cosmological parameters $\theta_{i, j} = \{\Omega_{\rm c},\sigma_{8}\}$. We assume a Gaussian likelihood and a cosmology-independent covariance matrix \cite{Carron2013,Darsh2019} computed at the fiducial cosmology ($\theta_{\rm fid} = \{0.264,0.811\}$). Assuming flat priors and dropping the model-dependence, the posterior distribution is proportional to the likelihood, $L$:

\begin{equation}
    p(\theta|\textbf{x})\propto L(\theta)\propto \mathrm{exp}\left( -\frac{1}{2}\Delta\textbf{x}^{T}\hat{C}^{-1}_{\rm fid}\Delta \textbf{x}\right).
\end{equation}
Here, $\textbf{x}$ is our observable (i.e., each tSZ statistic), $\Delta\textbf{x}$ is the difference between the mean statistic predicted at a cosmology within the prior bounds and its fiducial value:
\begin{equation}
    \Delta\textbf{x}=\bar{\textbf{x}}-\bar{\textbf{x}}_{\rm fid}
\end{equation}
and $\hat{C}^{-1}_{\rm fid}$ is the inverse covariance matrix:  
\begin{gather}
    C_{\rm fid}=\frac{1}{N_{\rm fid}-1}\sum_{i=1}^{N_{\rm fid}}(\textbf{x}_{i}-\bar{\textbf{x}})(\textbf{x}_{i}-\bar{\textbf{x}})^{T}.\\
\end{gather}
The hat $\hat{}$ denotes that the covariance has been re-scaled following Ref.~\cite{Hartlap2007} to remove bias in its estimation
\begin{equation}
    \hat{C}_{\rm fid}^{-1}=\frac{N_{\rm fid}-d-2}{N_{\rm fid}-1}C_{\rm fid}^{-1},
\end{equation}
where $d$ is the size of the observable vector $\Delta\textbf{x}$ and $N_{\rm fid}=34,560$ is the number of simulations used to compute the covariance. In this work, $d$ ranges between $1-300$, depending on the exact set of statistics used to obtain the constraints and the binning choices. In all cases, the correction factor is very small (e.g., for the fiducial noiseless case when all of the descriptors are included $d=157$ and the correction factor is $\sim0.995$). We leave the exploration of using a cosmology-dependent covariance or a non-Gaussian form of the likelihood to future work. 

Since our inference is in a 2D parameter space, we can easily compute the likelihood directly on a fine grid of the parameter values within the prior bounds. In this work, we use $1000 \times 1000$ linearly spaced points between $\Omega_{c}\in\{0.21,0.32\}$ and $\sigma_{8}\in\{0.6,1\}$. At each point, we interpolate $\bar{\mathbf{x}}$ using a 2D Clough-Tocher cubic interpolator \cite{scipy}. We construct the interpolator using the mean summary statistics, $\bar{\mathbf{x}}$, that have been computed over $N_{\rm sims}=5,112$ tSZ maps at each of the 153 cosmologies in the suite and $N_{\rm fid}=34,560$ at the fiducial. We can then determine the likelihood at each of the 10$^{6}$ points in the grid and obtain the confidence contours. We use the area of the $68\%$ confidence contours, $A_{68}$, as the figure of merit in this work. We also compare the marginalized constraints on $\sigma_{8}$. For this, we compute the mean parameters as 
\begin{equation}
    \theta_{i}^{\rm mean}=\frac{\int L(\theta_{i}, \theta_{j})\theta_{i} d\theta_{i} d\theta_{j}}{\int L(\theta_{i}, \theta_{j})d\theta_{i}d\theta_{j}},
\end{equation}
and quote the errors by choosing the parameter values closest to the specified confidence limit of the marginalized likelihood.

\section{Results: Noiseless Maps}\label{sec:noiseless}
\subsection{Cosmological dependence}
\begin{figure*}
    \centering
    \includegraphics[width=\textwidth]{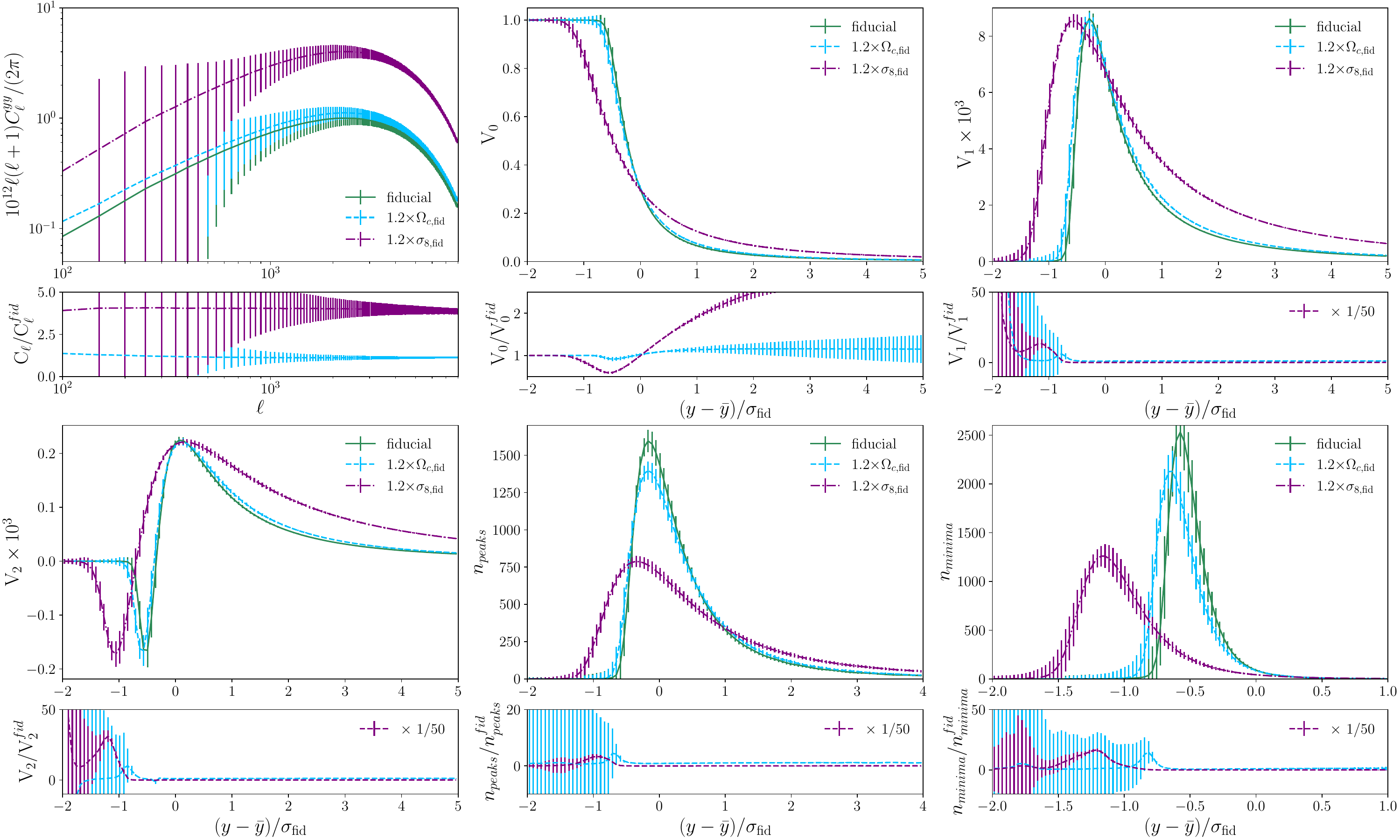}
    \caption{Summary statistics measured for three example cosmologies: the fiducial (green solid) cosmology and two cosmologies with $\Omega_{c}$ (blue dashed) and $\sigma_{8}$ (purple dot-dashed) increased by $20\%$. From left to right, the plots correspond to the power spectrum, three MFs ($V_{0}$, $V_{1}$, $V_{2}$), peaks, and minima. The stronger dependence on $\sigma_{8}$ is clearly evident. The bottom panels show the ratio between each of the two cosmologies and the fiducial. Variance on each of the observables is also shown.}
    \label{fig:3cosmo}
\end{figure*}

In Fig.~\ref{fig:3cosmo} we show the summary statistics computed for three example cosmologies over 5,112 maps: the fiducial and two others with $\Omega_{c}$ and $\sigma_{8}$ increased by 20$\%$ to illustrate their dependence on cosmological parameters. For the four moments, we list their values in Table \ref{tab:example_moments}. For the moments, we compute their values over 34,560 maps for the fiducial cosmology in order to reduce their sample variance and determine accurate scalings with cosmological parameters.

First, we can clearly see the much stronger dependence of the observables on $\sigma_{8}$ compared to $\Omega_{c}$.\footnote{For clarity, we therefore scale some of the ratios in  Fig.~\ref{fig:3cosmo} by a factor of 50.}  In the case of the power spectrum, the cosmological dependence is predominantly in the amplitude, which scales as $\propto\sigma_{8}^{7.4-7.7}$, in contrast to $\propto\Omega_{c}^{0.6-1.7}$ across the different multipoles. Similar dependence on the cosmological parameters has previously been shown, for example in  Refs.~\cite{KomatsuSeljak2002,HillPajer2013,Bolliet2018}.  Ref.~\cite{Bolliet2018} found the power spectrum to scale as $\propto\sigma_{8}^{8.1}\Omega_{m}^{3.2}$. 

\begin{figure*}
    \centering
    \includegraphics[width=0.7\textwidth]{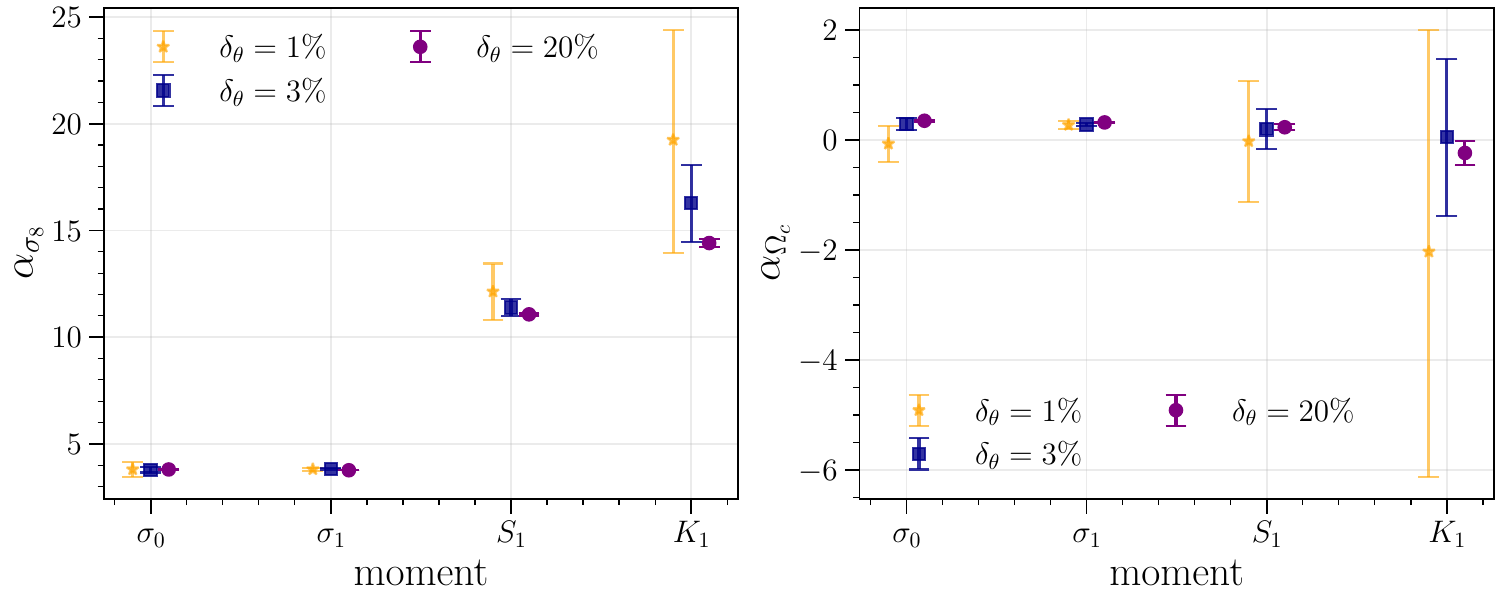}
    \caption{The left (right) panel shows the $\sigma_{8}$ ($\Omega_{c}$) parameter scalings for the four moments computed using different sets of simulations: from Fisher simulations with parameters increased by $1$ or $3\%$ and from the example simulations, where parameters are increased by $20\%$. The errors are obtained by considering the lowest and highest scaling within the errors for a given moment.}
    \label{fig:scalings}
\end{figure*}

For the four moments, we find scalings ranging from $\propto\sigma_{8}^{3.8-14.4}$ and $\propto \Omega_{c}^{(-0.2)-(+0.4)}$, where the range in the powers reflects the different dependence for the four moments between $\sigma_{0}$ and $K_{1}$ as listed in Table~\ref{tab:example_moments}.  As expected, the higher-order moments scale more steeply with parameters than the lower-order moments. Since $\sigma_0$ is the rms, the scalings for the variance, $\sigma_{0}^2$, are similar to those found in Ref.~\cite{HillSherwin2013} (i.e., $\sigma_0^2 \propto \sigma_8^{7.6}$ scales the same as the power spectrum, as expected). We note that $K_{1}$ scales negatively with $\Omega_{c}$. A possible explanation for such dependence could be that in order to have a fixed $\sigma_{8}$ while increasing $\Omega_{c}$ (with $h$ fixed), the amplitude of the primordial power spectrum, $A_{\rm s}$, needs to decrease. The decrease in $A_{\rm s}$ leads to a reduction in the halo mass function that is larger than the amplification caused by the increase in $\Omega_{c}$, which can lead to such negative scaling for some of the moments. For example, we have $A_{\rm s}=2.04\times10^{-9}$ at the fiducial cosmology with $\Omega_{c}=0.264$ and $\sigma_{8}=0.811$, and $A_{\rm s}=1.64\times10^{-9}$ at the cosmology with $\Omega_{c}=0.3168$ and $\sigma_{8}=0.811$. While we find that the mean values of the moments are converged within the standard error as a function of $N_{\rm sims}$, the exact parameter scalings can vary due to some variation within the error bars. Therefore, we list the scalings computed using these example cosmologies and also show the scalings for each of the moments computed using different sets of simulations in Fig.~\ref{fig:scalings}. The error bars here are obtained by taking the value of each moment at the fiducial and the example cosmology that leads to the smallest and largest scaling within the error of the given moment (e.g., in the case of positive moments, the lowest (highest) limit on the scaling can be obtained by taking the highest (lowest) fiducial value for the moment and the lowest (highest) value for the moment at a given cosmology).

We find that $V_{0}$ shows a noticeably different dependence on the two parameters, which leads to the degeneracy breaking that we discuss in subsequent sections. In particular, we see that $\Omega_{c}$ leads to a moderate, mostly constant increase in $V_{0}$ at high thresholds and a `dip' for a narrow range of low-significance thresholds. We find that $\sigma_{8}$ has the largest effect on the high-significance thresholds as well as a larger `dip' across a wider range of low-significance thresholds. This is similar to what has been found in Refs.~\cite{Hill2014PDF, Thiele2019} for the one-point PDF. $V_{1}$ and $V_{2}$ both exhibit a more pronounced cosmological dependence at the intermediate threshold values, as well as an overall steeper dependence at the high-significance thresholds. This suggests their higher sensitivity to $\Omega_{c}$ and $\sigma_{8}$. As in the case of $V_{0}$, the two cosmological parameters have different effects on $V_{1}$ and $V_{2}$.

Finally, $\Omega_{c}$ and $\sigma_{8}$ affect different bins of the peak and minima histograms. We note that the peak histograms are generally more extended, covering a wider range of $y$ values, while the minima histograms are narrower. The average number of peaks for the three cosmologies is $\approx 27,000-29,000$, which is $\approx240-260$ peaks/deg$^{2}$. For the minima counts, the total number is of the same order:  $\approx 26,000-27,000$ or  $\approx240$ minima/deg$^{2}$. 

\begin{table*}[]
    \centering
    \begin{tabular}{|c|c|c|c|c|}
    \colrule
    & $\sigma_{0}\times10^{-6}$ &  $\sigma_{1}\times10^{-7}$ & $S_{1}\times10^{-19}$ & $K_{1}\times10^{-23}$ \\
    \colrule
    fiducial& 1.54 &  1.27& $-3.22$& $-1.94$\\
    
    $1.2\times\sigma_{8}$ &3.08 & 2.52&
 $-24.2$ & $-26.8$\\
  $\alpha_{\sigma_{8}}$& 3.8&3.8&11&14\\
    $1.2\times\Omega_{c}$& 1.64&  1.34& $-3.35$&  $-1.86$\\
    $\alpha_{\Omega_{c}}$&0.4&0.3&0.2&$-0.2$\\
    \colrule
    \end{tabular}
    \caption{Values and approximate cosmological parameter scalings for the four moments. The values are listed for the fiducial cosmology and two cosmologies with $1.2\times\sigma_{8, \rm fid}$ and $1.2\times\Omega_{c, \rm fid}$. The power-law scalings are defined as $\propto\sigma_{8}^{\alpha_{\sigma_{8}}}$ and $\propto\Omega_{c}^{\alpha_{\Omega_{c}}}$.}
    \label{tab:example_moments}
\end{table*}

\subsection{Constraints}
\begin{figure*}[h]
    \includegraphics[width=\linewidth]{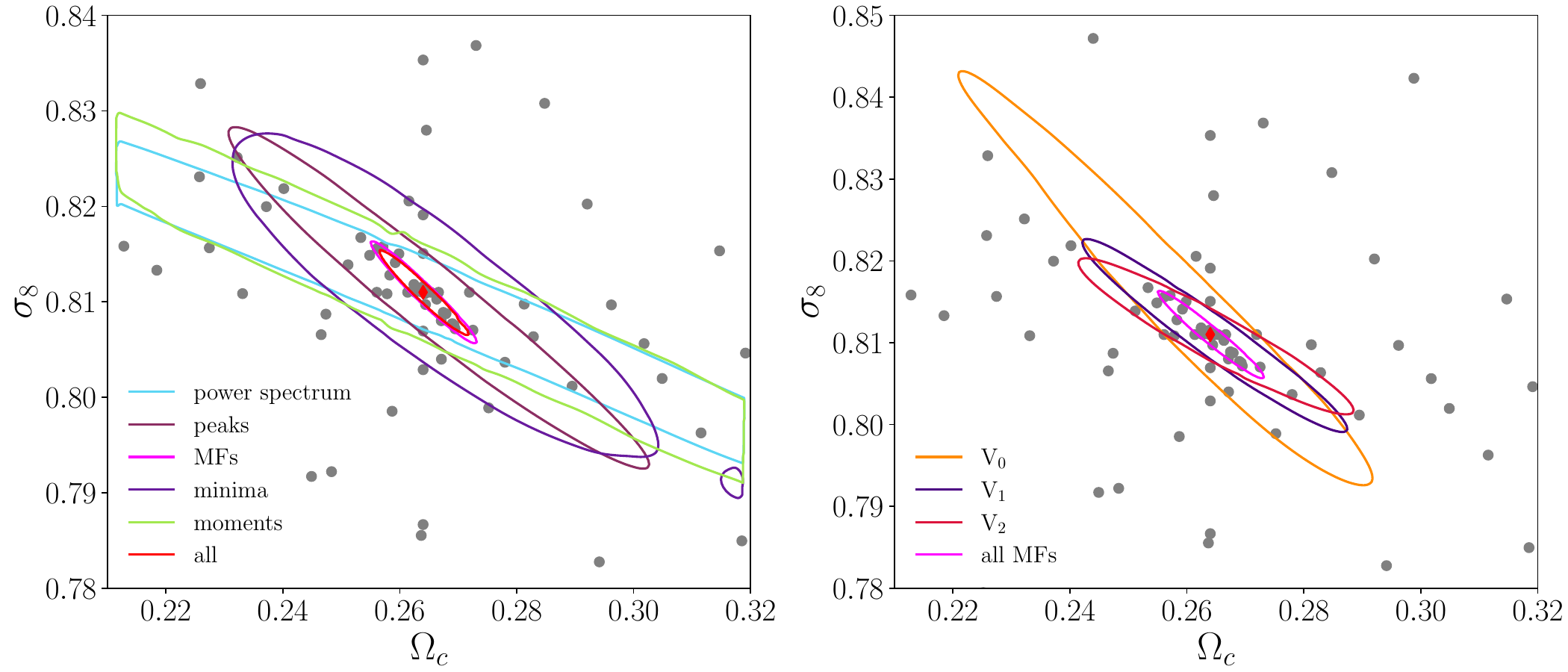}
    \caption{\textit{Left:} 68$\%$ credible contours for the \emph{noiseless} maps (smoothed with a Gaussian kernel with $\theta_{\rm FWHM}=1.4$ arcmin). The MFs yield $21\times$, and all statistics combined yield $29\times$, improvement over the power spectrum, as quantified via the area of the $1\sigma$ contours. \textit{Right:} Results obtained with each of the MFs separately. The improved constraints are driven by the combination of the three MFs, and dominated by $V_{1}$ and $V_{2}$ roughly equally.}
    \label{fig:noiseless_all}
\end{figure*}

In Fig.~\ref{fig:noiseless_all} (left), we show the credible contours obtained using noiseless tSZ maps with the power spectrum, Minkowski functionals, peaks, minima, and moments. Note that we do not re-scale the likelihood to a realistic $f_{\rm sky}$ in this section and only focus on the relative constraining power for each of the descriptors.\footnote{For example, if we assume that the constraints would approximately scale by the square root of the sky area, the errors are expected to be $\sim 12$ times tighter for $f_{\rm sky}=0.4$.} We quantify the constraining power in terms of the 1$\sigma$ contour area ($A_{68}$), the comparison to the power spectrum  ($A_{68}^{\rm ps}/A_{68}$), and the marginalized constraints on $\sigma_{8}$ ($\hat{\sigma}_8$). As can be seen in Fig.~\ref{fig:noiseless_all}, the power spectrum does not have enough sensitivity to break the degeneracy between $\sigma_{8}$ and $\Omega_{c}$, and thus the 1$\sigma$ contour area for the power spectrum can be considered a lower bound.

We find that when all the observables are combined, the constraints are $\sim29$ times tighter than from the power spectrum alone. The MFs yield tighter constraints than any of the other observables, while breaking the $\Omega_{c}-\sigma_{8}$ degeneracy. Compared to the power spectrum, the contour area is $21$ times smaller. In fact, the constraints from all five statistics combined are only $1.4$ times tighter than from the MFs alone. 

Peak counts are the second-most constraining statistic. Peaks similarly break the $\Omega_{c}-\sigma_{8}$ degeneracy, but with overall wider credible contours in both the $\Omega_{c}$ and $\sigma_{8}$ directions, which yields $1.6$ times tighter constraints than the power spectrum. Since the constraints from all the observables are tighter than from using the MFs alone, this suggests that peaks and MFs are complementary but still probe some of the same information.

Minima counts also break the degeneracy and have a similar tilt to the peak counts, but are less constraining in the direction of the best-constrained parameter combination. The contour area is actually larger than for the power spectrum ($A_{68}^{\rm ps}/A_{68}=0.84$). Since minima counts trace the `valleys' of the tSZ field, it makes intuitive sense why this statistic may contain less information in comparison to the peaks, for example.

Finally, we find that the four moments also achieve worse results compared to the power spectrum, $A_{68}^{\rm ps}/A_{68}=0.75$, and do not break the degeneracy between the two cosmological parameters. This is possibly due to the fact that the moments do not contain all the scale information captured by the power spectrum. The moments of gradients contain some spatial/scale information but likely not all of it. Additional scale information could further be included by combining moments measured for different smoothing scales, as we explore for the MFs in \S\ref{sec:noisy}.

While the most striking effect of using MFs or combining all the summary statistics is the breaking of the $\Omega_{c} - \sigma_{8}$ degeneracy and a substantial shrinkage in the contour area, the marginalized errors on $\sigma_{8}$ are also considerably improved. With MFs, the error on $\sigma_{8}$ is almost $3$ times smaller than the error obtained using the power spectrum. On the other hand, all other observables give $\sigma_{8}$ errors comparable to that from the power spectrum. 

\begin{figure*}
    \centering
    \includegraphics[width=\textwidth]{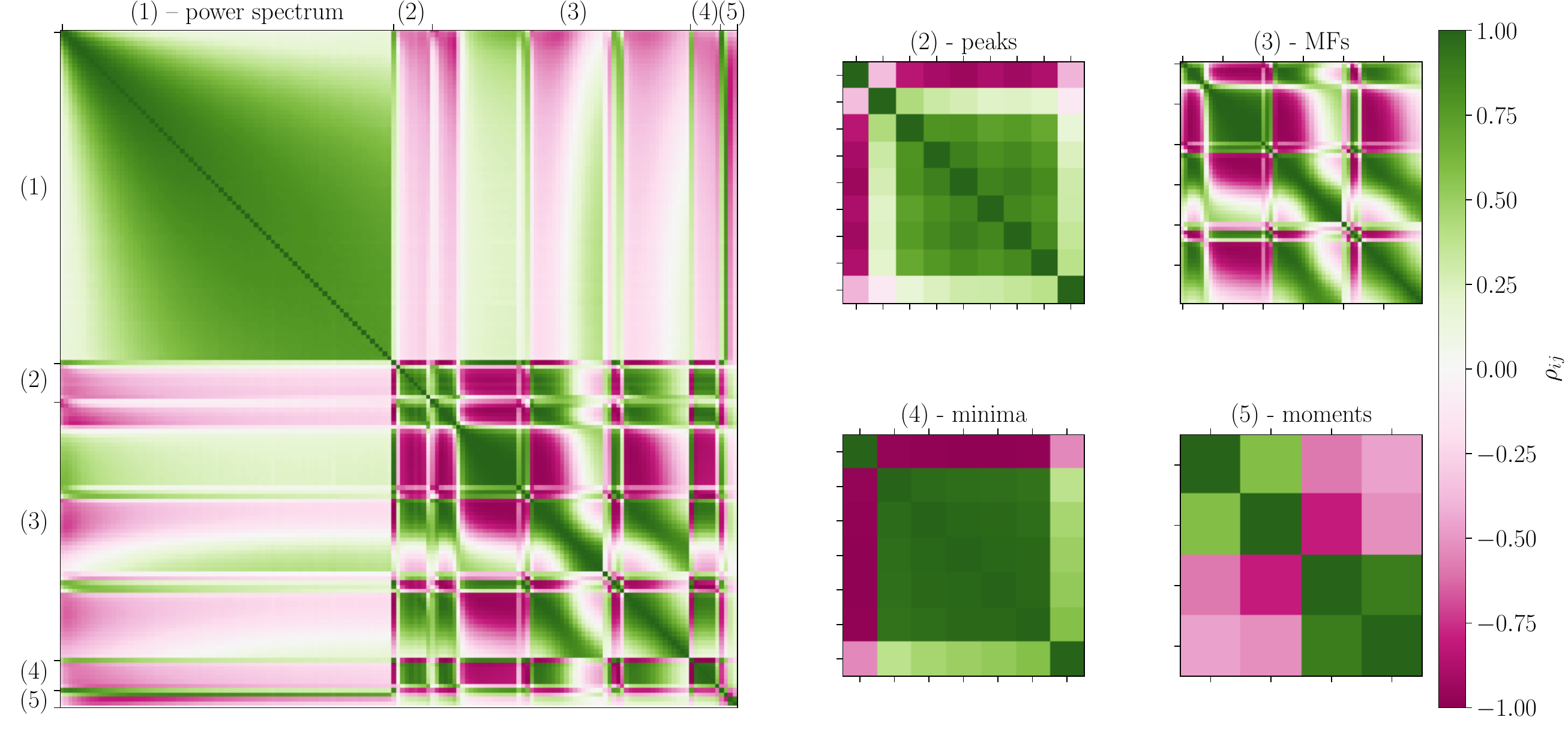}
    \caption{Correlation matrix for the \emph{noiseless} observables. The elements correspond to (1) power spectrum (77 bins), (2) peaks (9 bins), (3) MFs (60 bins), (4) minima (7 bins), and (5) moments (4 values).  For the latter four observables, the correlation matrices are also shown separately on the right for clarity. All observables exhibit significant correlation across neighboring bins. The MFs show relatively low cross-correlation between the three MFs, and $V_{1}$ and $V_{2}$ have relatively low auto-correlation amongst their own bins. The cross-correlations between the different summary statistics are generally lower than the auto-covariances within each statistic.}
    \label{fig:corr}
\end{figure*}

Looking at the constraints from each of the observables combined with the power spectrum, we see that using the power spectrum and  MFs 
results in 23 times improvement over the power spectrum alone (as compared to the $21\times$ improvement found using the MFs alone, compared to the power spectrum). This suggests that the MFs do not contain all the information captured by the power spectrum and the two observables could be combined for better constraints. It is possible that the MFs are not sensitive to some of the angular scales that we measure with the power spectrum. We find similar results for peaks, minima, and moments, which when combined with the power spectrum give $\sim 3.4, 1.9,$ and $1.2$ times tighter constraints compared to using the power spectrum alone, respectively. This shows that each of these summary statistics is complementary to the power spectrum, probing different information in the tSZ maps. In particular, we note that the constraints from using the peaks and power spectrum are much tighter than using peaks alone. Similarly, combining minima counts and moments with the power spectrum achieves much better results than when each of those statistics is used individually. In general, this suggests that using the power spectrum in combination with one of these additional observables, in particular MFs or peak counts, achieves the best results. 

In Fig.~\ref{fig:corr} (left), we show the correlation matrix for all the observables combined, defined as $\rho_{ij}=C_{ij}/\sqrt{C_{ii}C_{jj}}$, where $C$ is the covariance. For clarity, we also show the correlation matrices for peaks, MFs, minima, and moments separately in the right panel. We can see the strong correlation across most multipoles for the power spectrum. Only the lowest bin has weak correlation with the rest. For the peaks, intermediate thresholds show the most correlation, while the second and the last bin contain independent information. Similar dependence is observed for the minima, except the correlations (and anti-correlations) are much stronger across the bins. $V_{0}$ shows strong correlations across the different bins. As expected, similar results were shown in Ref.~\cite{Hill2014PDF} for the one-point PDF. As previously suggested, this is likely due to the same clusters contributing to multiple Compton-$y$ bins, as is most likely the case for the peaks and minima counts as well. $V_{1}$ and $V_{2}$ show less correlation between further-separated bins. The correlation is also low for some of the cross-elements between the three MFs. Finally, the moments show overall low correlation with one another. While there is significant correlation for the different bins for most of the observables, the cross-covariance between the different summary statistics, in particular between the power spectrum and the other descriptors, is mostly low. This is consistent with the results discussed above and the tight constraints that can be achieved by combining the various observables.

We summarize the $1\sigma$ contour areas for each of the statistics and the marginalized errors on $\sigma_{8}$ in Table~\ref{tab:noiseless_constraints}. We do not include the errors on $\Omega_{c}$ because the $\sigma_8 - \Omega_c$ degeneracy is not broken in all cases, including for the power spectrum, which serves as our baseline statistic. 

\begin{table}[t]
\caption{Constraints obtained from different summary statistics on \emph{noiseless} maps ($10.5\times10.5$ deg$^{2}$ and smoothed with a Gaussian kernel $\thetagauss=1.4 $ arcmin): $68\%$ contour area ($A_{68}$), comparison to the power spectrum area ($A_{68}^{\rm ps}$/$A_{68}$), and the mean value and marginalized error on $\sigma_{8}$ ($\hat{\sigma_{8}}$). We highlight the most significant improvement in the constraining power over the power spectrum in boldface. \label{tab:noiseless_constraints}}
\begin{tabular}{|l|c|c|c|}
\colrule
Observable & $A_{68}$ $\times10^{4}$ & $A_{68}^{\rm ps}$/$A_{68}$ & $\hat{\sigma_{8}}$\\
\colrule
power spectrum & 7.82&1.0&$0.810_{-0.009}^{+0.01}$\\
peaks & 4.92&1.6&$0.810\pm0.012$\\
MFs &0.367&\textbf{21}&$0.811_{-0.003}^{+0.004}$\\
minima &9.31&0.84&$0.811\pm0.011$\\
moments &10.4&0.75&$0.810\pm0.011$\\
all combined &0.267&\textbf{29}&$0.811\pm0.003$\\
\colrule
power spectrum + peaks & 2.28&3.4&$0.810\pm0.012$\\
power spectrum + MFs &0.346&\textbf{23}&$0.811_{-0.003}^{+0.004}$\\
power spectrum + minima &4.11&1.9&$0.811\pm0.011$\\
power spectrum + moments &6.73&1.2&$0.810\pm0.011$\\
\colrule
$V_{0}$&6.21&1.3&$0.816_{-0.019}^{+0.013}$\\
$V_{1}$&2.16&3.6&$0.810\pm0.008$\\
$V_{2}$&2.32&3.4&$0.811\pm0.006$\\
\colrule
\end{tabular}
\end{table}

\subsection{Convergence with respect to the number of simulations}\label{sec:noiseless_convergence}

\begin{figure}
    \centering
    \includegraphics[width=\linewidth]{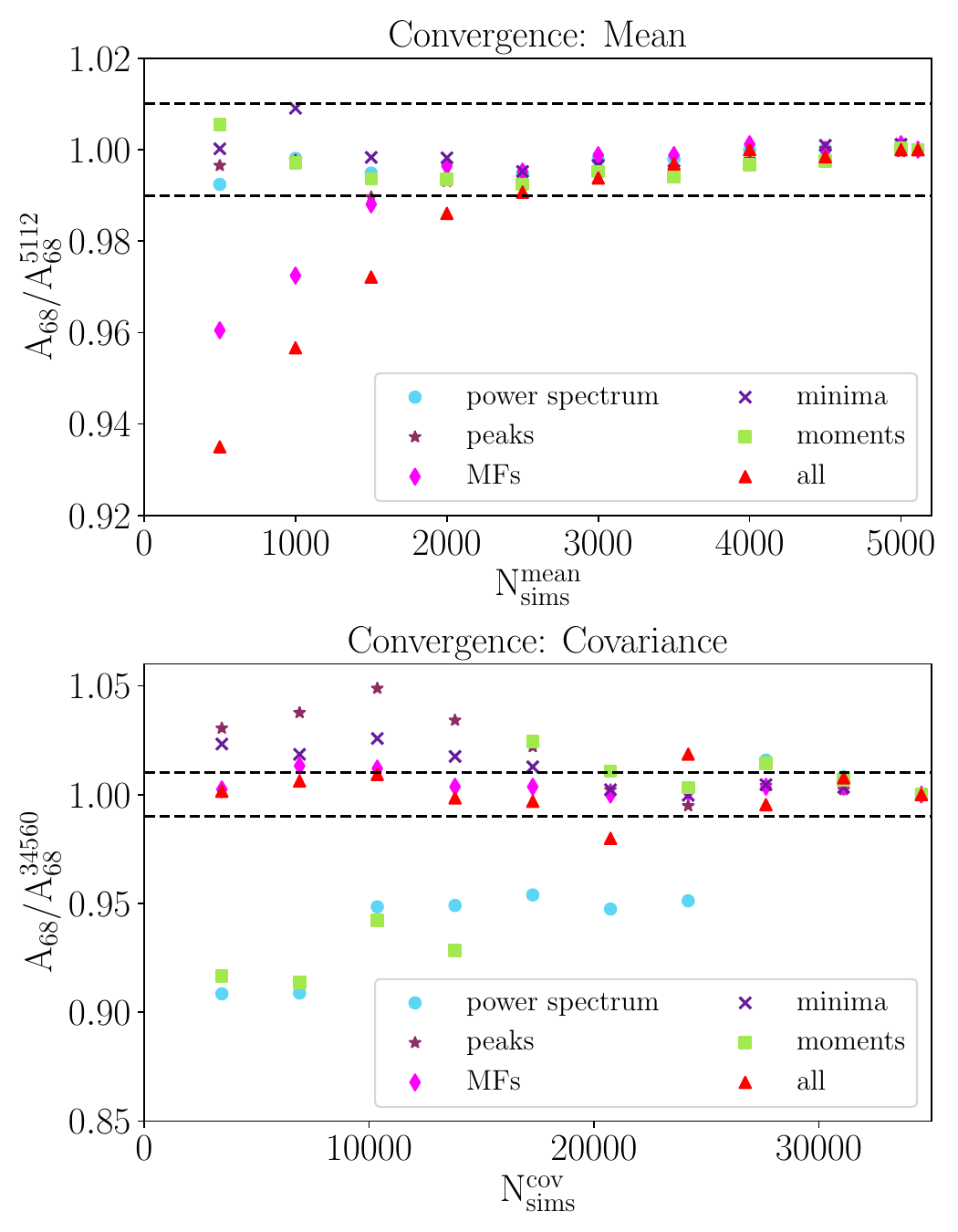}
    \caption{Convergence of the constraints with respect to the number of simulations. We show the ratio of the $68\%$ confidence area obtained using a subset of simulations to the constraints obtained using all available simulations. \textit{Top:} Convergence of the mean summary statistic. \textit{Bottom:} Convergence of the covariance. Horizontal dashed lines mark $\pm1\%$.}
    \label{fig:nsims_area}
\end{figure}

We check that we have a sufficient number of realizations for the covariance estimation by computing the contours using half of the simulations at the fiducial cosmology ($N_{\rm fid}=17,280$ for both the covariance and the mean summary statistic at the fiducial cosmology). With this number of simulations, the confidence area from all the summary statistics combined remains the same ($<1\%$ difference). Note that some of the other summary statistics require a larger number of simulations based on the individual covariance convergence tests described below (Fig. \ref{fig:nsims_area}). However, this initial check shows that the constraints from all summary statistics combined, which give smooth contours, are stable to the number of simulations used at the fiducial cosmology to $\sim$percent-level. 

Next, we check that we have simulated enough realizations at each of the other cosmologies (i.e., that the mean statistics across the suite are converged). Although we have determined this via convergence tests performed prior to running the suite of simulations (see Appendix~\ref{sec:convergence}), we can check this directly by computing the credible contour areas using a different number of simulations to compute the mean summary statistic at each cosmology (while keeping the number of simulations for the covariance estimation at $N_{\rm fid}=34,560$). For these tests, we compute the constraints using all of the summary statistics combined and each summary statistic individually. We plot the ratio between the 1$\sigma$ area computed using all $N_{\rm sim}=5,112$ realizations at each cosmology and an area computed using some subset of the realizations in the upper panel of Fig.~\ref{fig:nsims_area}. We see that for all the observables, the contours are converged to within $\sim1\%$ at 2,500 realizations and to within $\sim2\%$ at 2,000 realizations. 

Similarly, we can explicitly check the convergence as a function of the number of simulations used for the covariance estimation (while using 5,112 maps for the mean, including the fiducial cosmology). We see that the covariance is converged for each of the summary statistics (lower panel of Fig.~\ref{fig:nsims_area}). For MFs, as few as $\approx3,500$ are needed. 

\subsection{Minkowski functionals}
To determine if the constraining power is driven by any specific MFs, we look at the credible contours for each of the MFs separately, which are shown in Fig.~\ref{fig:noiseless_all} (right). $V_{0}$, $V_{1}$, and $V_{2}$ all have distinct degeneracy directions. Thus, the tightest constraints are achieved when all three are combined. Conceptually, $V_{0}$ is equivalent to the one-point PDF, while $V_{1}$ and $V_{2}$ probe spatial information beyond what is captured in the `area'. It has been previously proposed that the one-point PDF is a near-optimal cosmological statistic for the tSZ field, since the tSZ field is dominated by the one-halo term and the one-point PDF captures information from all zero-lag moments up to arbitrarily high order~\cite{Hill2014PDF,Thiele2019}. However, here we see that the $1\sigma$ contour area is roughly $2.9\times$ and $2.7\times$ smaller for $V_{1}$ and $V_{2}$, respectively, and $17\times$ smaller when all three MFs are combined, compared to using $V_{0}$ alone. One interpretation of these results is that each of the MFs are sensitive to a different range of halo masses, which leads to the different degeneracy directions for each of the MFs.

\begin{figure*}
    \centering
    \includegraphics[width=\textwidth]{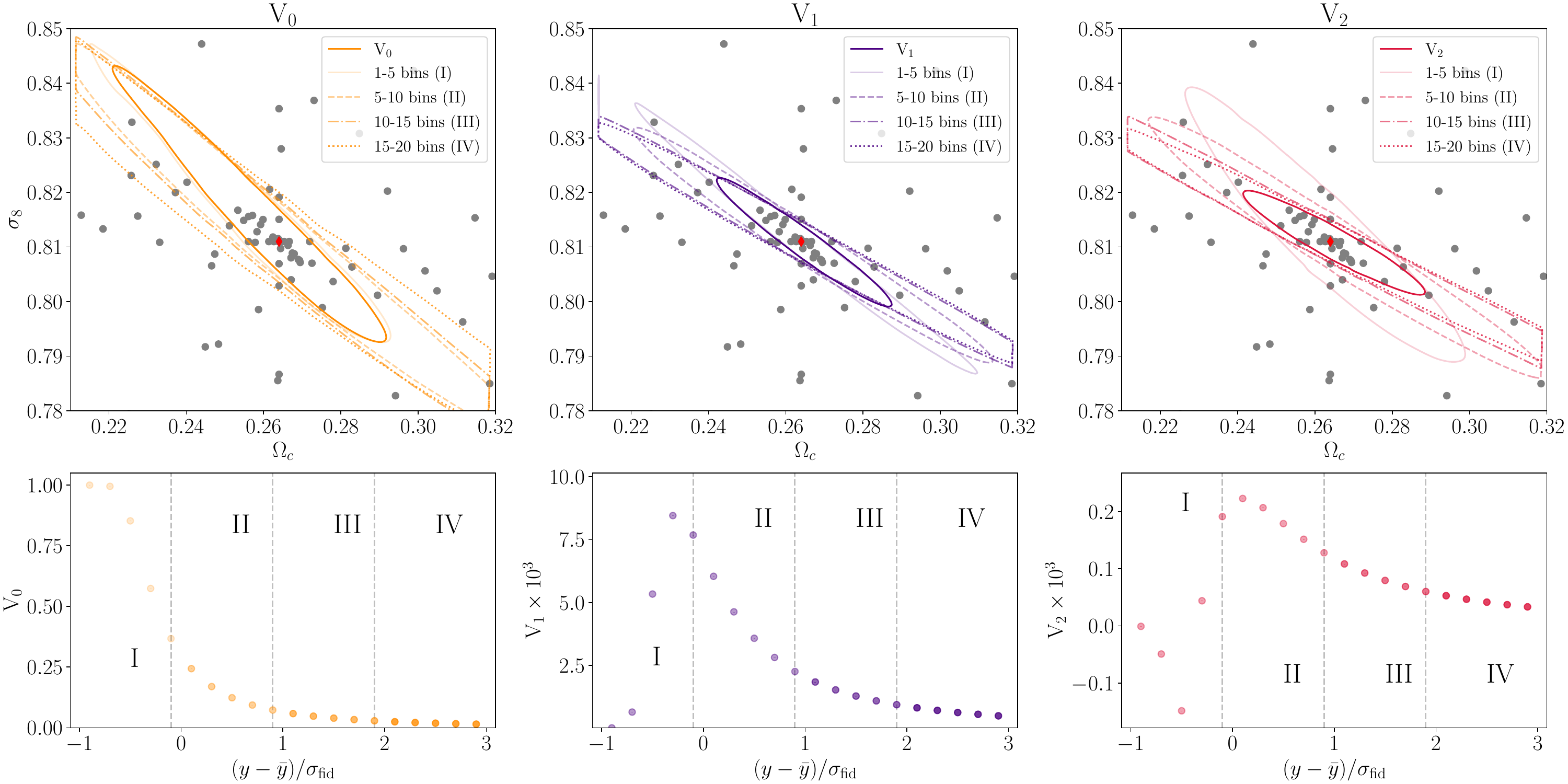}
    \caption{\textit{Upper:} Constraints from each of the MFs using a subset of the thresholds over which each is measured (the fiducial threshold bins are between $\nu\in\{-1\sigma_{\rm fid},3\sigma_{\rm fid}\}$). For $V_{0}$, almost all of the constraining power is in the five lowest significance thresholds, while for $V_{1}$ and $V_{2}$, it is in the lowest half of the bins.  \textit{Lower:} Each of the MFs at the fiducial cosmology. The vertical (dashed) lines and the shades of the markers correspond to the intervals considered for the constraints in the upper panel.}
    \label{fig:cosmo_sens_MF}
\end{figure*}

Next, we look at which MF bins contain the most sensitivity to cosmology. To do this, we compute credible intervals using only a subset of the thresholds over which we have measured the MFs (recall that we use 20 linearly-spaced bins between thresholds $\nu\in\{-1\sigma_{\rm fid},3\sigma_{\rm fid}$\}). In the case of $V_{0}$, most of the constraining power lies in the lowest significance bins. In Fig.~\ref{fig:cosmo_sens_MF}, we show the constraints computed for different subsets of the thresholds, from lowest to highest values, and show that the constraints are almost entirely coming from the lowest five bins (e.g., the area of the 1$\sigma$ contour is just $\sim20\%$ larger than when using all the thresholds). This can be explained by the large sample variance of the high threshold bins. Even though the tail of $V_{0}$ contains significant cosmological information as seen in Fig.~\ref{fig:3cosmo}, the sample variance reduces its constraining power. Note also how the degeneracy direction of the ellipse is different for each of the threshold bin subsets, suggesting that different parts of the halo mass function are probed.

In the case of  $V_{1}$ and $V_{2}$, the lowest half of the bins contains most of the sensitivity to cosmology. For these MFs, the full shape is important in capturing the information (e.g., the first 10 of the thresholds), while much less information lies in the high-significance tails. This is most likely due to the noticeably different degeneracy directions between the lowest bins and the rest (see Fig.~\ref{fig:cosmo_sens_MF}). On the other hand, the  intermediate and the high threshold bins seem to be probing similar information. From a practical standpoint, this means that we do not necessarily need to measure the MFs at the highest threholds. We can thus avoid possible numerical noise that can arise from the low values in those bins. Moreover, since the sample variance is the most difficult to capture at the variance levels corresponding to the most rare, massive clusters, possibly fewer simulations would be needed to obtain stable results with respect to the number of simulations and \verb|hmpdf| halo mass function discretization settings when the threshold range is even more limited (e.g., to $\nu\in\{\nu_{\rm min},\nu_{\rm max}\}\in\{-\sigma_{\rm fid},\sigma_{\rm fid}\}$).

\subsection{Comparison to Fisher forecast}
In our analysis, we directly compute the posteriors for the cosmological parameters through the likelihood computation on a fine grid. This is possible via the suite of simulations that allow us to build an interpolator and predict the value of the summary statistics at any other cosmology. We compare our results to constraints obtained using the Fisher matrix formalism in order to: 
\begin{itemize}
    \item [(1)] Further cross-check that our results from the suite are converged. 
    \item [(2)] Determine if Fisher forecasts may be sufficient for predicting the constraints from the MFs (given how close the constraints from the MFs are to an ellipse).
\end{itemize} 

The Fisher matrix $F_{ij}$ for the parameters $\theta_{i,j}$, and a likelihood $L$, can be written as follows
\begin{equation}
    F_{ij}=\left\langle\frac{\partial \ln L}{\partial\theta_{i}}\frac{\partial \ln L}{\partial\theta_{j}}\right\rangle.
\end{equation}
The marginalized errors can then be estimated via
\begin{equation}
\begin{split}
    \sigma_{i}=\sqrt{\left(F^{-1}\right)_{ii}}
\end{split}
\end{equation}
and the 68$\%$ Fisher ellipse area using
\begin{equation}
    A_{\rm Fisher}=\pi\Delta\chi^{2}ab \,,
\end{equation}
where $a$ and $b$ can be computed using Eq.~\eqref{eq:ellipse_params} and $\Delta\chi^{2}\approx2.3$ for 68$\%$ C.L. 
More precisely, to compute the Fisher matrix from simulations assuming a Gaussian likelihood, we need the derivatives of the mean summary statistic vectors ($\mathbf{\bar{x}}$) with respect to parameters $\theta_{i,j}$ and the inverse covariance matrix computed at the fiducial cosmology ($\hat{C}_{\rm fid}^{-1}$) using $N_{\rm fid} = 34,560$ simulations and re-scaled with the Hartlap factor, as previously defined in \S\ref{sec:inference}. The Fisher matrix can then be calculated using  
\begin{equation}
    F_{ij}=\frac{\partial{\mathbf{\bar{x}}}}{\partial\theta_{i}}\hat{C}_{\rm fid}^{-1}\frac{\partial{\mathbf{\bar{x}}}}{\partial\theta_{j}}.
\end{equation}

To determine the derivatives, we consider three set-ups, where $\Omega_{c}$ and $\sigma_{8}$ are perturbed one at a time by $0.5\%, 1\%,$ and $3\%$ from their fiducial values (a total of 12 cosmologies in addition to the fiducial). We calculate the derivatives from these simulations using a central difference scheme:
\begin{equation}
    \frac{\partial\mathbf{\bar{x}}}{\partial\theta_{i}}\approx\frac{\mathbf{\bar{x}}\rvert_{\theta=\theta_{i}+\delta\theta_{i}}-\mathbf{\bar{x}}\rvert_{\theta=\theta_{i}-\delta\theta_{i}}}{2\delta\theta_{i}}
\end{equation}

\begin{figure*}
    \centering
        \includegraphics[width=\textwidth]{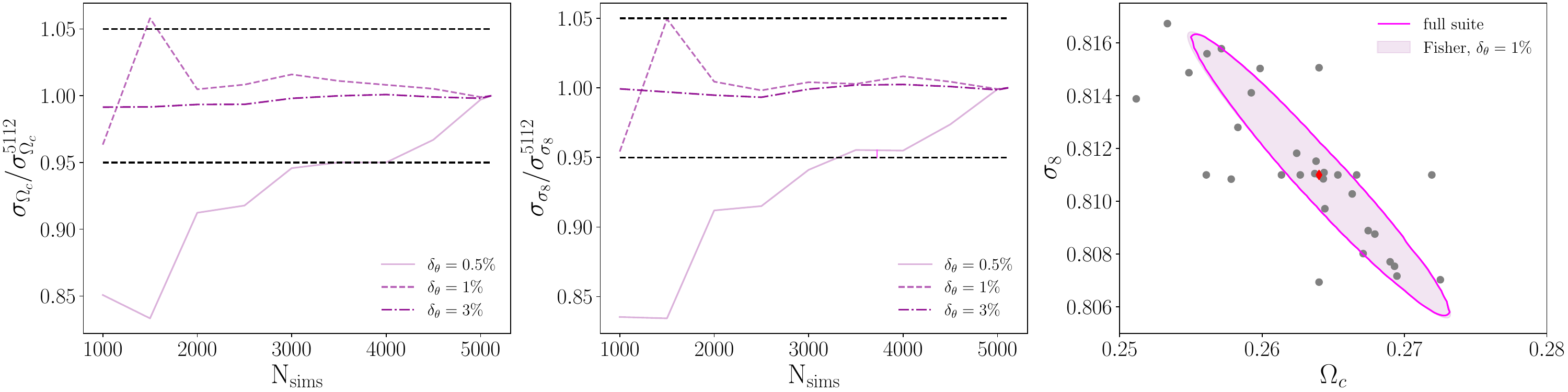}
        \caption{Comparison of the constraints from MFs obtained using the Fisher formalism versus using the full suite of simulations to compute the confidence contours as described in \S\ref{sec:inference}. From left to right, panels (1) and (2) show the convergence of the Fisher errors on $\Omega_{c}$ and $\sigma_{8}$ with respect to the number of simulations used to compute the derivatives: the ratio between the error computed using $N_{\rm sims}$ as indicated and all 5,112 simulations. The convergence is shown using Fisher cosmologies perturbed by $0.5\%,1\%,$ and $3\%$. Horizontal black dashed lines mark $\pm5\%$. Figure (3) shows the Fisher ellipse (computed for $\delta_{\theta}=1\%$) and the credible contours computed using the full suite. The contour areas are in agreement to $\approx2\%$. }\label{fig:fisher_compare}
\end{figure*}

We compute the Fisher forecasts for the noiseless case, both for the MFs and the power spectra. It has been shown in previous works that simulation-based Fisher forecasts can give overly confident parameter error bars if the derivatives are not numerically converged, i.e., an insufficient number of simulations have been used~\cite{Coulton2023fisher, Coulton2023png}. To check for numerical convergence, we plot the $1\sigma$ errors as a function of the number of simulations in Fig.~\ref{fig:fisher_compare} for the MFs. We see good convergence for the $1\%$ and $3\%$ cases -- the Fisher error bars are stable with respect to the number of simulations used to compute the derivatives. In the case of MFs, the Fisher constraints are also in very good agreement with the results obtained from using the full suite. In Fig.~\ref{fig:fisher_compare}, we show the Fisher forecast ellipse using the $1\%$-variation cosmologies and the contours computed from the entire suite. The contour areas agree to $\approx2\%$ from the two methods. While the marginalized errors are numerically converged for the $3\%$-variation simulations, the Fisher forecast using these simulations gives overly optimistic constraints (ellipse area is smaller by $\approx 15\%$ compared to the constraints computed using the full suite), suggesting the derivatives computed using these simulations are not accurate. As an additional numerical convergence check for the $1\%$-variation Fisher simulations, we split the simulations used to compute the derivatives into subsets of 1,700 and 3,412 simulations and compute the Fisher forecast using these two sets of simulations. From this comparison, we conclude that the contours computed from the full suite are consistent with Fisher forecasts. This suggests that in future work, we can use Fisher forecasts for the MF-based forecasts instead of generating a full simulation suite if, for example, we want to compute the forecasts using more expensive simulations such as hydrodynamical simulations or $N$-body simulations with painted pressure profiles.

For the power spectrum case, we do not see the same numerical convergence with the number of simulations. One possibility is that the power spectrum has much less sensitivity to $\Omega_{c}$, as is evident in the shape of the contours and the degeneracy in the $\Omega_{c}-\sigma_{8}$ direction. We experiment with discarding some of the power spectrum bins that lead to the noisiest derivatives but that does not improve the convergence of the Fisher errors. We compute the forecast for $\sigma_{8}$ alone and find good convergence both as a function of the number of simulations and across all $\delta\theta=0.5,1,3\%$. The 1$\sigma$ unmarginalized error on $\sigma_{8}$ is $\sigma_{\sigma_{8}}=0.004$, consistent with the suite-based results.

\subsection{Mass range}

In addition to the statistical approach, the tSZ effect can be used to constrain cosmological parameters via cluster counts. It is therefore interesting to determine whether the tight constraints obtained via MFs are primarily driven by halos within a similar range of masses as those that would be detected in tSZ cluster surveys or by less massive halos, below the detection threshold. A careful comparison between cluster count cosmology analyses and the tSZ MFs, which would incorporate the subtleties regarding selection functions for cluster surveys, and the marginalization over unknown astrophysics (e.g., modeling the mass-observable relation for clusters or varying the pressure profile parameters in the tSZ map simulations) is outside the scope of this paper. Instead, we perform a simple test to see if the MF-based constraints are predominantly sourced by the most massive halos in our simulated maps. 

\begin{figure*}
\centering
    \begin{minipage}{0.4\linewidth}
    \centering
        \includegraphics[width=\textwidth]{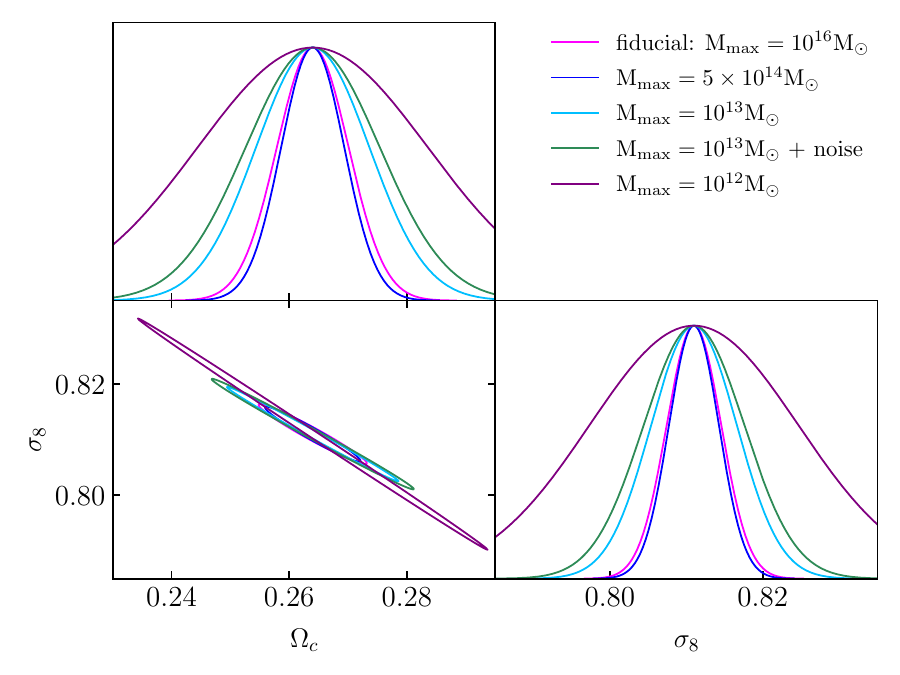}
    \end{minipage}
    \begin{minipage}{0.5\linewidth}
    \centering
        \includegraphics[width=\textwidth]{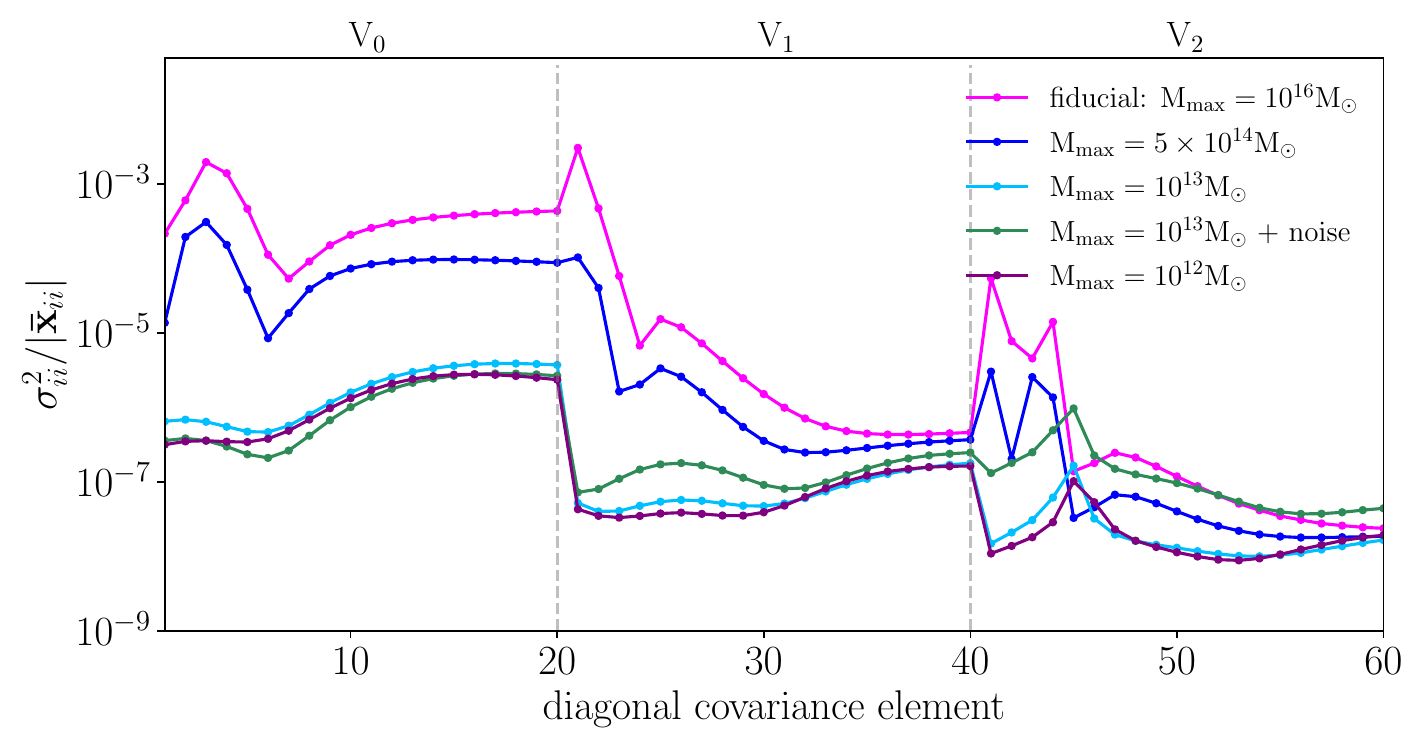}
    \end{minipage}
    \caption{\textit{Left:} Fisher forecast constraints computed using simulations with a different maximum halo mass. The constraints tighten when the most massive halos are excluded (dark green). Lowering the maximum mass to 10$^{13}$ ${\rm M}_{\odot}$ (cyan) weakens the 1D marginalized constraints, but does not affect the total Fisher ellipse area. \textit{Right:} Variance for each of the MF bins computed at the fiducial cosmology using simulations with the different mass cuts. The variance is reduced as the maximum mass is lowered, since the sample variance is highest for the most massive, rare halos.}
    \label{fig:fisher_mass}
\end{figure*}

To do this, we generate several sets of tSZ maps for which we restrict the maximum halo mass to a different value than the fiducial $M_{\rm max}=10^{16}$ ${\rm M}_{\odot}$ and check how significantly the forecasts are impacted.\footnote{We choose to set a different $M_{\rm max}$ rather than computing the forecast for the most massive halos directly (i.e., setting $M_{\rm min}$ to a higher value) to avoid computing MFs over maps with a large number of zero-value pixels.} Instead of simulating several additional full suites, we focus on a Fisher forecast comparison. As we have shown in the previous section, a Fisher forecast using MFs is in good agreement with our results from the full suite. Therefore, we choose to simulate tSZ maps for only five cosmologies: the fiducial model and four cosmologies with $\Omega_{c}$, $\sigma_{8}$ perturbed by $\pm1\%$ one at a time (see \S\ref{sec:sims} for the exact $\Omega_{c}$, $\sigma_{8}$ values). 

We set the upper mass limit to $M_{\rm max}=5\times10^{14} \,\, {\rm M}_{\odot}$, motivated by the minimum mass in the latest ACT cluster catalogue from Ref.~\cite{Hilton2021}. As with the rest of the forecasts in this paper, we use 34,560 random realizations at the fiducial cosmology and 5,112 realizations at the perturbed cosmologies. In the upper panel of Fig.~\ref{fig:fisher_mass}, we compare the Fisher constraints obtained using the full mass range in our fiducial analysis to the one with the reduced upper limit. We find that the constraints do not degrade by excluding the most massive halos. In fact, the Fisher errors somewhat tighten (the Fisher ellipse area is smaller by $\approx18\%$). We check the convergence of the results by splitting the simulations used to compute the derivatives in the Fisher forecast into two separate batches (1,700 and 3,412 realizations) and computing the forecasts using each of the sets separately. We find that the constraints are in good agreement from both sets of maps, which means that we have a sufficient number of simulations to compute the derivatives. Similarly, we split the simulations at the fiducial cosmology in half and obtain the constraints using each separate half of the simulations for the covariance estimation. We find that our results are consistent.

As additional tests, we compute Fisher errors using additional sets of maps with even more aggressive upper mass cuts of $M_{\rm max}=10^{13}$ ${\rm M}_{\odot}$ and $10^{12}$ ${\rm M}_{\odot}$, which we also include in Fig.~\ref{fig:fisher_mass}. In the $M_{\rm max}=10^{13}$ ${\rm M}_{\odot}$ case, we see that the marginalized constraints for $(\Omega_{\rm c}$, $\sigma_{8})$ weaken by $\approx (58\%, 57\%)$, respectively. The total contour area, however, remains roughly the same ($<1\%$ change). The Fisher ellipse stretches out in the $\Omega_{\rm c}$ direction but tightens along the $\sigma_{8}$ direction. To cross-check this result, we compute the constraints adding a small amount of noise to the maps (white noise 0.01 $\mu$K-arcmin with no filters applied) as, realistically, the faint tSZ signal from low-mass halos will be washed out by noise. In this case, the overall constraints are indeed worse than those with $M_{\rm max}=10^{16}$ ${\rm M}_{\odot}$. For $M_{\rm max}=10^{12}$ ${\rm M}_{\odot}$, we lose roughly half of the constraining power and the errors on $(\Omega_{\rm c}$, $\sigma_{8})$ become larger by a factor of $\approx (3.2, 3.8)$. We note that the convergence of the Fisher errors requires more simulations for the cases with lower maximum halo mass cuts. We still confirm that the size of the errors converges to within $\approx5\%$ as a function of the number of simulations used for computing the derivatives (i.e., the convergence test shown in Fig.~\ref{fig:fisher_compare}). 

We speculate that the improvement in the constraints when using $M_{\rm max}=5\times10^{14}$ ${\rm M}_{\odot}$ is due to a decrease in the sample variance when removing the most massive objects in the maps. To show this, we plot the variances for each of the MF bins in the right panel of Fig.~\ref{fig:fisher_mass}. It can be clearly seen that the MF variance decreases as we lower the maximum halo mass. As expected, we observe that the variance is very similar for the $M_{\rm max}=10^{13}$ ${\rm M}_{\odot}$ and $10^{12}$ ${\rm M}_{\odot}$ cutoffs. We show the full absolute covariance matrices for the four maximum mass cases in Appendix~\ref{app:cutoffs} (see Fig.~\ref{fig:mass_corr}), which similarly show a reduction in the amplitude of the off-diagonal terms when using lower maximum mass cutoffs. We also include the correlation matrices, which show less correlation across the different MF bins as the halo mass upper limit is lowered.

These Fisher forecast tests with different maximum halo mass bounds have two main takeaways:
\begin{enumerate}
    \item[(1)] The constraints from MFs are driven primarily by the halo masses below the mass range that is individually detected in tSZ cluster surveys. In particular, the constraints improve by $\approx1.2\times$ when we set the maximum mass to $M_{\rm max}=5\times10^{14}$ ${\rm M}_{\odot}$, and remain roughly the same when we set $M_{\rm max}=10^{13}$ ${\rm M}_{\odot}$. This suggests that the masses driving the MF constraints are lower than those that are directly detected for cluster cosmology analyses. Joint analyses using tSZ MFs and cluster counts, therefore, have the potential to achieve even tighter constraints on the cosmological parameters, since the two observables probe different halo mass ranges and can thus be highly complementary to each other. This conclusion is consistent with what has been learned about the tSZ power spectrum. For example, Ref.~\cite{HurierLacasa2017} has found that there is a lack of correlation between the high-$\ell$ tSZ power spectrum and cluster counts, as the two observables predominantly probe clusters at different redshift ranges.
    \item[(2)] Removing the most massive halos improves the constraints that can be achieved with the tSZ MFs by reducing the sample covariance, since the most massive objects have the highest sample variance. This again is consistent with the results found for the tSZ power spectrum (e.g., see Refs.~\cite{Shaw2009, HillPajer2013, Bolliet2020, Osato2021, Rotti2021}).
\end{enumerate}
We summarize the results from these comparisons in Table~\ref{tab:mass_range}. For each of the constraints, we measure the MFs using 20 bins between $\nu\in\{-\sigma_{\rm fid}^{x}, 3\sigma_{\rm fid}^{x}\}$ with the following variances: $\sigma_{\rm fid}^{5\times10^{14}}=1.1\times10^{-6}, \sigma_{\rm fid}^{10^{13}}=8.2\times10^{-8}, \sigma_{\rm fid}^{10^{13}+\rm{noise}}=9.1\times10^{-8}, \sigma_{\rm fid}^{10^{12}}=1.1\times10^{-8}$.

\begin{table}[]
    \centering
    \begin{tabular}{|c|c|c|c|c|}
    \colrule
    $M_{\rm max}$ [${\rm M}_{\odot}$] & $A_{\rm F}\times10^{5}$ & $A_{\rm F}^{16}$/$A_{\rm F}$&$\sigma_{ \sigma_{8}}$&$\sigma_{\Omega_{c}}$\\
    \colrule
    $10^{16}$ &3.75&1.0&$0.004$&$0.006$\\
    $5\times10^{14}$ &3.08&1.2&$0.003$&$0.005$\\
    $10^{13}$ &3.76&1.0&$0.006$&$0.01$\\
    $10^{13}$ + noise &5.63&0.7&$0.007$&$0.011$\\
    $10^{12}$ &7.42&0.5&$0.014$&$0.02$\\
    \colrule
    \end{tabular}
    \caption{We compare the Fisher forecast constraints from MFs using simulations with different mass ranges. We list the maximum halo mass, $A_{\rm F}$ (the Fisher ellipse area), $A^{16}_{\rm F}$/$A_{\rm F}$ (comparison to the fiducial constraints with $M_{\rm max}=10^{16}$ ${\rm M}_{\odot}$), and the marginalized errors on $\Omega_{c}$ and $\sigma_{8}$. We set $M_{\rm min}=10^{11}$ ${\rm M}_{\odot}$ in all cases. For the noise in the fourth case, we use 0.01 $\mu$K-arcmin white noise and apply no filters to the maps. We find that in the idealized noiseless case, most of the constraining power of the MFs lies in the lower-mass halos, as the most massive objects contribute more to the variance than to the signal (see \S\ref{sec:noiseless}).}
    \label{tab:mass_range}
\end{table}

\section{Results: Including Noise}\label{sec:noisy}
\begin{figure*}
    \centering
    \includegraphics[width=\linewidth]{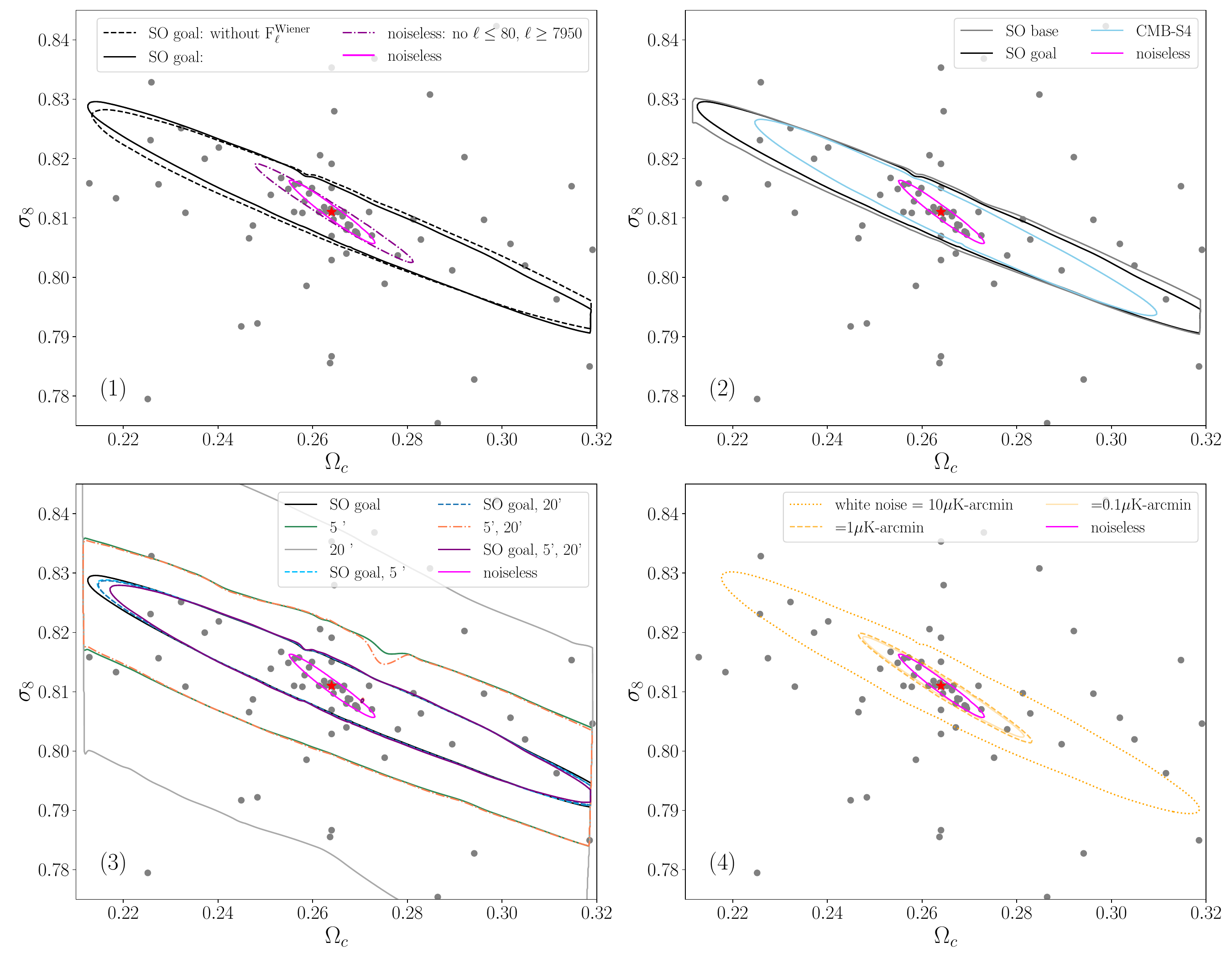}
    \caption{Constraints from MFs ($V_0$, $V_1$, and $V_2$) using various set-ups in comparison to the results for the noiseless maps (solid magenta in all panels).  We emphasize that these results are computed for $10.5 \times 10.5$ deg${}^2$ maps, not maps covering the expected survey footprints of upcoming experiments. The panels show, starting from top left: (1) Effects of low- and high-pass filtering on the noiseless constraints and the impact of the Wiener filter on the noisy constraints for the SO goal experimental set-up. (2) Constraints obtained using MFs on noisy maps for various experimental set-ups: SO baseline, SO goal, and CMB-S4. (3) Effects of combining MFs measured for different smoothing scales. (4) Constraints for various levels of white noise.}
    \label{fig:noisy_4}
\end{figure*}

\subsection{Minkowski functionals}
First, we study the impact of noise on the MFs, which are the most constraining summary statistic in the noiseless case. 
\subsubsection{Different experimental set-ups}

We show the effects on the constraints of applying additional filters to the maps in panel (1) of Fig.~\ref{fig:noisy_4}. As described in \S\ref{sec:sims}, we low- and high-pass filter the tSZ maps when combining with noise maps. We do this to avoid any problematic numerical behavior due to the high noise at the smallest scales and to avoid extrapolation beyond the $\ell$ range available in the public SO tSZ noise power spectra. Since we want the signal maps to contain the same angular scale range as the noise realizations, we apply the same filter to both the tSZ maps and the GRFs generated from the noise power spectra. As shown in panel (1) of Fig.~\ref{fig:noisy_4}, applying the filter increases the contour area by a factor of $\approx2.3$ due to the loss of modes. In the same plot, we also show the effects of applying a Fourier-space Wiener filter to noisy maps (see Eq.~\eqref{eq:wiener} in \S\ref{sec:sims}). To obtain the noisy constraints, we combine the smoothed tSZ maps with a GRF generated from the post-component-separation tSZ noise power spectra (more details are included in \S\ref{sec:sims}). We apply this filter to up-weight the scales where the signal is more dominant and down-weight the scales where the noise is large. For the SO goal noise levels, the Wiener filter improves the constraints by a factor of $\approx1.1$. For these contours, we compute the MFs using 30 bins between $\nu\in\{-3\sigma_{\rm fid}^{x},3\sigma_{\rm fid}^{x}\}$ with $\sigma_{\rm fid}^{\rm SO\,goal}=1.9\times10^{-6}$ and $\sigma_{\rm fid}^{\rm SO\,goal\, without\, F^{\rm Wiener}_{\ell}}=3.2\times10^{-6}$. Adding noise makes the maps more Gaussian and thus the measured MFs are more symmetric around the zero-value threshold (i.e., the tails span similar numbers of negative and positive-value threshold bins) and broader compared to the noiseless cases in Fig.~\ref{fig:3cosmo}. We therefore choose to extend the range of thresholds over which we measure the MFs in order to capture their full shape.

Next, in panel (2) of Fig.~\ref{fig:noisy_4}, we show the results for three possible experimental set-ups: SO baseline, SO goal, and CMB-S4 noise levels. In each case, we have applied a corresponding Wiener filter. For these constraints, we again compute the MFs using 30 bins between $\nu\in\{-3\sigma_{\rm fid}^{x},3\sigma_{\rm fid}^{x}\}$ with the following variances: $\sigma_{\rm fid}^{\rm SO, base}=2.1\times10^{-6}$, $\sigma_{\rm fid}^{\rm SO, goal}=1.9\times10^{-6}$, $\sigma_{\rm fid}^{\rm CMB-S4}=1.7\times10^{-6}$. With CMB-S4-level noise, the MF constraints are improved by $\approx1.6\times$ compared to the constraints using SO goal noise. The SO baseline noise level gives constraints that are $\approx0.9\times$ as tight compared to the SO goal noise level. These results are summarized in Table~\ref{tab:noise_diff_levels}.

\begin{table}[]
    \centering
    \begin{tabular}{|c|c|c|c|}
    \colrule
    Noise level & $A_{68}\times10^{4}$ & $A_{68}^{\rm SO goal}$/$A_{68}$&$\hat{\sigma_{8}}$\\
    \colrule
    SO baseline &9.18&0.9&$0.810\pm0.011$\\
    SO goal &7.92&1.0&$0.810\pm0.011$\\
    CMB-S4 &5.07&1.6&$0.810\pm0.011$\\
    \colrule
    \end{tabular}
    \caption{We compare the constraints from MFs for different experimental set-ups, as shown in panel (2) of Fig.~\ref{fig:noisy_4}. The results here are shown for $10.5\times10.5$ deg$^{2}$ maps that have been low- and high-pass filtered, smoothed, combined with noise maps and Wiener-filtered as described in \S\ref{sec:sims}. $A_{68}^{\rm SO goal}$ corresponds to the $68\%$ confidence area for the SO goal noise set-up to which we compare the SO baseline and CMB-S4 constraints.}
    \label{tab:noise_diff_levels}
\end{table}

\subsubsection{Smoothing scales}
It has been shown in previous works that combining different smoothing scales can tighten the constraints from MFs (e.g., ~\cite{Kratochvil2012MF}). Smoothing has the effect of adding some spatial or scale information, especially to $V_{0}$, which otherwise has none \cite{Kratochvil2012MF}. To test this on our noisy maps, we compute the MFs on two additional sets of noisy maps that have been additionally smoothed using a Gaussian kernel with $\thetagauss= 5$ arcmin and $\thetagauss= 20$ arcmin. Note that in this case we smooth the total `tSZ + noise' map. For these results,  we use the SO goal noise power spectra to generate the noise maps. We show the results in panel (3) of Fig.~\ref{fig:noisy_4}. 
The constraints are shown for:
\begin{itemize}
    \item tSZ + noise with a Wiener filter applied (denoted as SO goal, the same set-up as the one in panels (1) and (2) in Fig. \ref{fig:noisy_4} in black solid curve);
    \item tSZ + noise with $\thetagauss=5$ arcmin and $\thetagauss=20$ arcmin smoothing applied to the composite map (denoted as 5' and 20');
    \item constraints from combining the different cases.
\end{itemize}

We again compute the MFs using 30 bins between  $\nu\in\{-3\sigma_{\rm fid}^{x},3\sigma_{\rm fid}^{x}\}$ with $\sigma_{\rm fid}^{\rm SO,\,  5\,arcmin}=2.0\times10^{-6}$ and $\sigma_{\rm fid}^{\rm SO,\,  20\,arcmin}=0.9\times10^{-6}$. When we look at the two additional smoothing cases (5', 20') individually, simply smoothing the maps only weakens the constraints since we lose some of the small-scale information. The tightest constraints are from the fiducial case where simply a Wiener filter is applied. We find that combining the Wiener filter case with the 5 arcmin or 20 arcmin cases does not in any notable way improve the constraints over the Wiener filter constraints. In practice, combining scales increases the total dimension of the observable data vector and as such, the size of the covariance matrix and the number of simulations needed to properly estimate it. Our current results suggest that the gains from combining with 5 arcmin or 20 arcmin smoothing scales may not be large enough to be worth pursuing, but it may be worth testing smaller smoothing scales in future work. The 68$\%$ confidence areas for the different smoothing cases are listed in Table~\ref{tab:smoothing_results}.

The tests using different smoothing scales indicate that much of the information captured by the MFs is lost if the maps are smoothed with a Gaussian kernel at $\thetagauss\gtrsim5$ arcmin. In future works, it will be useful to explore additional smoothing scales, or different filters to determine which scales the MFs are most sensitive to.
\begin{table}[]
    \centering
    \begin{tabular}{|c|c|c|c|}
    \colrule
    Smoothing scale(s) & $A_{68}\times10^{4}$ & $A_{68}^{\rm SO goal}$/$A_{68}$&$\hat{\sigma_{8}}$\\
    \colrule
    SO goal & 7.92 & 1.0&$0.810\pm0.011$\\
    5' & 22.6 & 0.35&$0.809_{-0.014}^{+0.015}$\\
    20' & 60.6 &0.13&$0.804_{-0.027}^{+0.036}$\\
    SO goal, 5' & 7.82&1.01&$0.810\pm0.011$\\
    SO goal, 20' & 7.83&1.01&$0.810\pm0.011$\\
    5', 20' & 22.4 &0.35&$0.810_{-0.014}^{+0.015}$\\
    SO goal, 5', 20' & 7.70 &1.03&$0.810\pm0.011$\\
    \colrule
    \end{tabular}
    \caption{Constraints from MFs for the SO goal noise set-up using different smoothing scales, as described in \S\ref{sec:noisy} and shown in panel (3) of Fig.~\ref{fig:noisy_4}. The results here are shown for tSZ maps that are low- and high-pass filtered, smoothed with $\thetagauss=1.4$ arcmin, combined with noise maps and either Wiener filtered or additionally smoothed using  $\thetagauss=5$ or $20$ arcmin. The bottom four rows correspond to cases in which MFs measured from several setups are combined. We compare all the results to $A_{68}^{\rm SO\, goal}$, which corresponds to constraints from noisy maps that only include the effects of the $\thetagauss=1.4$ arcmin beam and a Wiener filter (row 1).}
    \label{tab:smoothing_results}
\end{table}

\subsubsection{White noise}
Next, we compute the constraints for three different white noise levels and show the results in the last panel of Fig.~\ref{fig:noisy_4}. We perform this exercise to determine the approximate noise levels at which we recover close to noiseless results. We use flat (white) noise power spectra, as opposed to realistic tSZ post-component-separation noise, to be able to tune the exact noise amplitude in a simple way. For these tests, we generate the noise maps assuming a flat power spectrum (recall that the tSZ signal maps are already beam-convolved with a FWHM = 1.4 arcmin beam) and convert them from $\mu$K-arcmin to $y$-rad units at 150 GHz, as described in \S\ref{sec:sims}. We low- and high-pass filter the maps and apply a Wiener filter. For the $0.1$ and 1 $\mu$K-arcmin noise levels, we compute the MFs using 20 bins between $\nu\in\{-1.5\sigma_{\rm fid}^{x}, 2.5\sigma_{\rm fid}^{x}\}$, where $\sigma_{\rm fid}^{0.1}=\sigma_{\rm fid}^{1}=1.5\times10^{-6}$. For the $10$ $\mu$K-arcmin noise level, we compute the MFs using 20 bins between $\nu\in\{-2\sigma_{\rm fid}^{10}, 2\sigma_{\rm fid}^{10}\}$ with $\sigma_{\rm fid}^{10}=1.2\times10^{-6}$. Again, we have chosen these ranges to capture the shape of the MFs since the maps are different for these three cases, as can be clearly seen in Fig.~\ref{fig:white_noise}. 

The credible contours in panel (4) of Fig.~\ref{fig:noisy_4} show that the constraints are converging toward the results from the noiseless maps as the noise level decreases. This can also be visually seen in Fig.~\ref{fig:white_noise} and from the fact that the variance $\sigma_{\rm fid}^{0.1}=\sigma_{\rm fid}^{1}=1.5\times10^{-6}$ is similar to the noiseless case. The areas of the 68\% contours are $0.883\times10^{-4}$, $1.14\times10^{-4}$, and $8.82\times10^{-4}$ for the 0.1, 1, and 10 $\mu$K noise levels, respectively. The ratios between the $68\%$ contour areas obtained using MFs computed for the noiseless maps and the different white noise levels are 0.4, 0.3, 0.04 (also listed in Table~\ref{tab:white_noise_results}). For the 0.1 and 1 $\mu$K-arcmin noise levels, the contour areas are larger only by $\approx3\%$ and $\approx33\%$ compared to the noiseless case where a low- and high-pass filter has been applied (see panel (1) in Fig.~\ref{fig:noisy_4}). This suggests that for upcoming experiments such as CMB-S4, which will have $\approx1$ $\mu$K-arcmin noise levels, improved component-separation techniques (e.g., via combination with CCAT-p data~\cite{Choi2020}) can push the constraints close to the noiseless scenario. 

\begin{table}[]
    \centering
    \begin{tabular}{|c|c|c|c|}
    \colrule
    noise [$\mu$K-arcmin] & $A_{68}\times10^{4}$ & $A_{68}^{\rm noiseless}$/$A_{68}$&$\hat{\sigma}_8$\\
    \colrule
    0.1&0.883&0.4&$0.811_{-0.005}^{+0.006}$\\
    1 &1.14&0.3&$0.811\pm0.006$\\
    10&8.82&0.04&$0.810_{-0.013}^{+0.012}$\\
    \colrule
    \end{tabular}
    
     \caption{Constraints from MFs for various white noise levels, as described in \S{\ref{sec:noisy}} and shown in panel (4) of Fig.~\ref{fig:noisy_4}. The results here are listed for $10.5\times10.5$ deg$^{2}$ tSZ maps that have been smoothed and combined with white noise maps. We list the $68\%$ contour area ($A_{68}$), comparison to noiseless constraints ($A_{68}^{\rm noiseless}/A_{68}$), and the mean value and marginalized error on $\sigma_{8}$ ($\hat{\sigma_{8}}$).}
    \label{tab:white_noise_results}
\end{table}

\subsection{All summary statistics and realistic survey area}

\begin{figure}
    \centering
    \includegraphics[width=\linewidth]{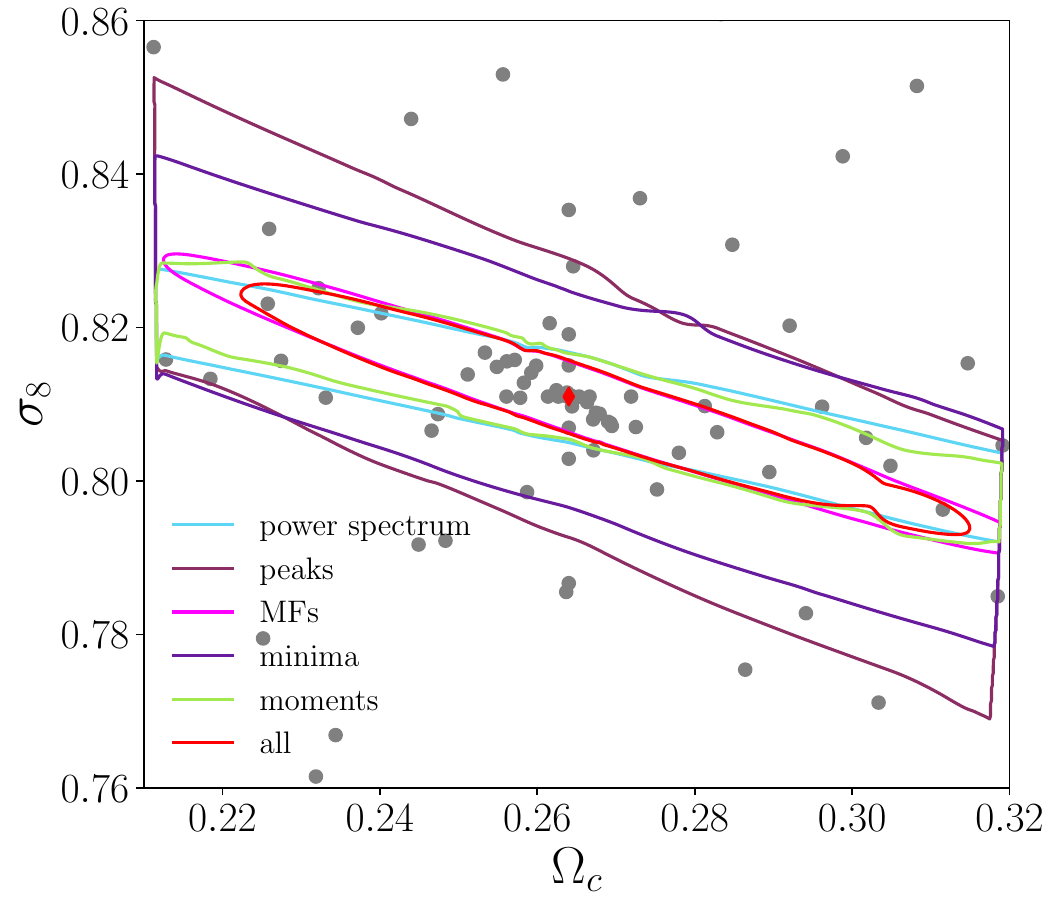}
    \caption{$68\%$ credible contours obtained using $10.5\times10.5$ deg$^{2}$ \emph{noisy} maps. Noise maps are generated using SO goal tSZ noise power spectra. The MFs give $1.6\times$ tighter constraints than the power spectrum. All summary statistics combined perform $1.8\times$ better than the power spectrum alone. The corresponding constraints are listed in Table~\ref{tab:noisy_constraints}.}
    \label{fig:noisy_main}
\end{figure}

We show the noisy results for all the summary statistics in the left panel of Fig.~\ref{fig:noisy_main}. These are computed using the SO goal post-component-separation tSZ noise power spectra and include the Wiener filter (i.e., the set-up matches the constraints shown for the MFs in the top two panels of Fig.~\ref{fig:noisy_4}, plotted in solid black). We find that in the noisy case, the MFs are still the most constraining observable ($\approx1.6\times$ tighter constraints than when using the power spectrum alone).  Combining the MFs and power spectrum yields $\approx 1.7\times$ improvement over the power spectrum alone. We find that the moments, peaks, and minima give worse constraints than the power spectrum. When combined with the power spectrum, peaks and minima result in $<10\%$ improvement over the power spectrum, while moments can achieve $1.2\times$ tighter constraints than from the power spectrum alone. When we combine all the statistics, we achieve a $\approx 1.8\times$ improvement over the power spectrum alone. We note that the contours from all summary statistics combined are somewhat jagged and the relative improvement in constraining power can be $\approx 1.7\times$ if fewer cosmologies in the suite are used to compute the likelihood surface. Nonetheless, we see that the constraints are primarily driven by power spectrum + MFs with slight improvement from the other statistics. As the tSZ map becomes more Gaussian once the noise is added, the MFs contain less additional information. As described in \S\ref{sec:descriptors}, MFs can be written in terms of the moments for a weakly non-Gaussian field. Since we find that the MFs still perform better than the four moments and better than the combination of the full set of nine moments as described in Appendix \ref{app:kurtosis}, it suggests that even in the presence SO-like levels of noise, the field is still highly non-Gaussian. This motivates a derivation of the analytic form of the MFs to higher-order corrections, which would allow determination of the exact set of moments needed to capture the information for the noisy tSZ maps, which could then be expressed analytically via, e.g., halo model tSZ polyspectra.

Examining each of the MFs individually, we find that $V_{1}$ and $V_{2}$ are notably more constraining than $V_{0}$ in the noisy case and that the combination of the MFs leads to the tightest results. In the noisy case, however, either using $V_{1}$ or $V_{2}$ results in constraints close to what can be achieved from combining all three ($\approx 1.4\times$ or $1.5\times$ versus $\approx 1.6 \times$ improvement over the power spectrum). We find that peaks, minima, and moments are not as constraining in the noisy case. This could possibly be due to noise `washing' out the peaks that are most sensitive to cosmology and Gaussianizing the field. Still, we find that combining the power spectrum with moments improves the constraints, suggesting that the two are complementary.

\begin{table}[t]
\caption{Constraints obtained from different summary statistics on \emph{noisy} maps ($10.5\times10.5$ deg$^{2}$): $68\%$ contour area ($A_{68}$) in the $\Omega_c - \sigma_8$ plane, comparison to the power spectrum contour area ($A_{68}^{\rm ps}$/$A_{68}$), and the mean value and marginalized error on $\sigma_{8}$ ($\hat{\sigma_{8}}$). For the noise maps, we use the SO goal tSZ noise power spectra, and apply both a low-/high-pass filter and a Wiener filter. \label{tab:noisy_constraints}}
\begin{tabular}{|l|c|c|c|}
\colrule
Observable & $A_{68}$ $\times10^{4}$ & $A_{68}^{\rm ps}$/$A_{68}$ & $\hat{\sigma}_8$\\
\colrule
power spectrum & 12.3&1.0&$0.810_{-0.009}^{+0.01}$\\
peaks & 38.6&0.32&$0.809\pm0.023$\\
MFs &7.92&1.6&$0.810\pm0.011$\\
minima &31.0&0.40&$0.809\pm0.018$\\
moments &12.2&1.01&$0.810_{-0.01}^{+0.011}$\\
all combined &6.78&1.8&$0.810_{-0.009}^{+0.01}$\\
\colrule
power spectrum + peaks & 11.7&1.1&$0.810\pm0.023$\\
power spectrum + MFs &7.23&1.7&$0.810\pm0.011$\\
power spectrum + minima &11.9&1.04&$0.810\pm0.018$\\
power spectrum + moments &10.6&1.2&$0.810_{-0.009}^{+0.01}$\\
\colrule
$V_{0}$&28.4&0.43&$0.810\pm0.019$\\
$V_{1}$&8.48&1.5&$0.810_{-0.013}^{+0.012}$\\
$V_{2}$&9.02&1.4&$0.811\pm0.012$\\
\colrule
\end{tabular}
\end{table}

\begin{figure*}
    \centering
    \includegraphics[width=\linewidth]{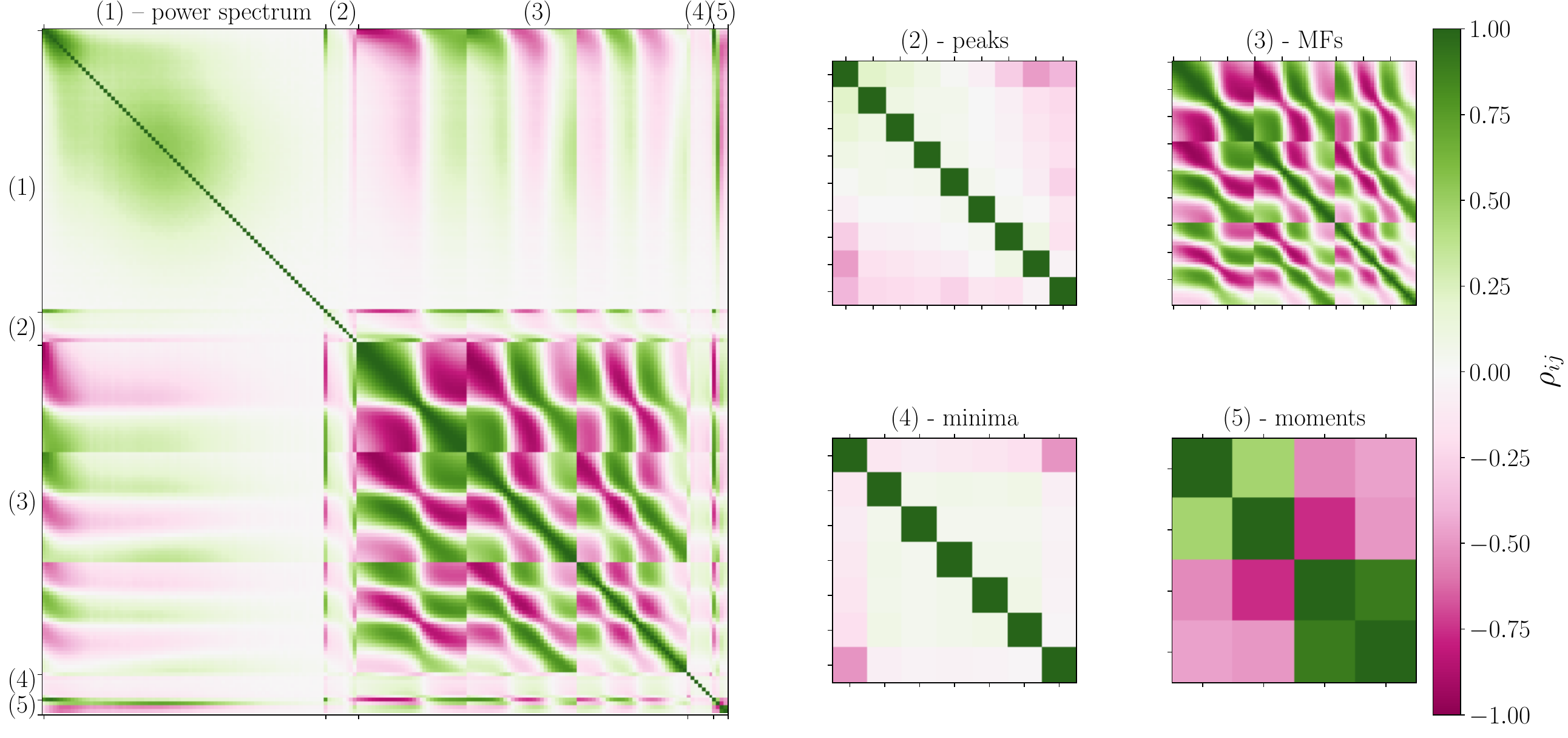}
    \caption{Correlation matrix for the tSZ statistics measured on \emph{noisy} maps (the associated parameter constraints are shown in Fig.~\ref{fig:noisy_main}). The elements correspond to (1) power spectrum (77 bins), (2) peaks (9 bins), (3) MFs (90 bins), (4) minima (7 bins), and (5) moments (four values). Noise reduces the correlation between the bins for the power spectrum, peaks, MFs, and minima, as well as the cross-covariance between the different observables.}
    \label{fig:noisy_corr}
\end{figure*}

We show the correlation matrix for the noisy maps in Fig.~\ref{fig:noisy_corr}. As expected, the noise significantly reduces the correlation across the different bins for most of the observables. We summarize the constraints for all the summary statistics in Table~\ref{tab:noisy_constraints}.

Finally, we scale the results from the power spectrum, MFs, and moments to a realistic sky fraction $f_{\rm sky}=0.4$ to approximately quantify the size of the parameter error bars that could be obtained from surveys like SO. Note that we do not include foregrounds in our simulated maps, apart from the residual foreground power in the post-component-separation tSZ noise power spectra, or marginalize over pressure profile parameters in this analysis. Therefore, these scaled forecasts represent an approximate idealized upper limit on how well we could constrain cosmological parameters with tSZ statistics. 

The noisy posteriors hit the prior bounds, as shown in Fig.~\ref{fig:noisy_main}. This makes scaling the marginalized errors on $\Omega_{c}$ and $\sigma_{8}$ computed directly from these posteriors not entirely accurate. Thus, instead, we compute the marginalized error on the best-constrained parameter combination $\Sigma_{8}=\sigma_{8}(\Omega_{c}/0.264)^{\alpha}$, where $\alpha$ is the slope of the line passing through the constraints, determined from the parameter covariance. The degeneracy directions vary for the observables: we find $\alpha=0.07$ and $0.11$ for the power spectrum and MFs, respectively. Therefore, it is best to compare how constraining each of the summary statistics is along its best-constrained parameter combination. We find $\Sigma_{8}^{\rm ps}=0.810_{-0.005}^{+0.006}$, and $\Sigma_{8}^{\rm MFs}=0.810\pm0.004$. Taking the average of the error bars (due to assymmetry), the MFs give  error bars tighter by $\approx 33\%$ than the power spectrum. Dividing the errors on $\Sigma_{8}$ by $\approx\sqrt{16501\mathrm{\, deg}^2/110\mathrm{\, deg}^2}\approx12.2$, we find that the MFs can constrain $\Sigma_{8}$ to $\approx 0.04\%$ precision, while the power spectrum alone can place an $\approx 0.06\%$ constraint.  We emphasize again that the proper inclusion of foreground contamination and marginalization over ICM astrophysics parameters will weaken these constraints, but nevertheless it is clear that the tSZ field contains significant cosmological information.

\section{Conclusions}\label{sec:discuss_conclusion}
In this work, we have studied the cosmological information content of several tSZ summary statistics: the power spectrum, MFs, peaks, minima, and moments. To do this, we have generated a large suite of simplified tSZ simulations (more than $800,000$ maps of size 10.5$\times$10.5 deg$^{2}$ each, with 0.1 arcmin pixels) with varying cosmological parameters $\Omega_{c}$ and $\sigma_{8}$. This is the first systematic study of several beyond-Gaussian tSZ statistics to our knowledge. Our main findings can be summarized as follows:
\newline
\newline
\textit{\underline{Noiseless maps}}
\begin{itemize}
    \item[(1)] MFs significantly outperform all the other summary statistics. Using MFs, we can achieve $\sim 21\times$ tighter constraints compared to those obtained from the tSZ power spectrum, as quantified by the area of the 68\% contour in the $\Omega_c-\sigma_8$ plane. The constraints are driven by the combination of the three MFs --- mostly by $V_{1}$ and $V_{2}$ --- due to their different degeneracies. Thus, we find that the MFs contain substantial additional information beyond the one-point PDF (i.e., since $V_{0}$ contains the information from the one-point function of the field). Combining all five summary statistics leads to a $\sim29$-fold improvement over the constraints from the power spectrum alone. Combining each of the four summary statistics with the power spectrum yields $\approx 23, 3.4, 1.9, 1.2\times$ improvement for MFs, peaks, minima, and moments, respectively, over using the power spectrum alone. 
    \item[(2)] We compute a Fisher forecast for several additional sets of simulations, which have a lower maximum halo mass cutoff: $M_{\rm max}=5\times10^{14},10^{13},10^{12}$ ${\rm M}_{\odot}$. We find that the MF-based constraints are improved by $\approx 1.2\times$ when we exclude the most massive halos, i.e., those that would be individually detected in tSZ cluster surveys ($M_{\rm max}=5\times10^{14}$ ${\rm M}_{\odot}$), while the marginalized errors on $\Omega_{\rm c}$ and $\sigma_{8}$ tighten by $\approx12\%$ and $8\%$, respectively. Such improvement in the constraints is likely due to a reduction in the sample variance from excluding the most massive halos. This suggests that combined analyses using tSZ MFs and cluster counts may provide powerful constraints.
\end{itemize}

\textit{\underline{Noisy maps}}
\newline
\begin{itemize}
    \item[(1)] In the noisy case, MFs again yield the best constraints, outperforming the tSZ power spectrum by a factor of $\sim 1.6$. Combining the five summary statistics gives a factor of $\sim 1.8$ improvement over the power spectrum alone. Combining each of the four summary statistics with the power spectrum gives $\approx 1.7, 1.2, 1.1, 1.04\times$ tighter constraints for MFs, moments, peaks, and minima, respectively, over using the power spectrum alone. We note that recently Ref.~\cite{Novaes2024} has similarly shown that MFs achieve the tightest constraints compared to peaks, minima, and the one-point PDF, when applied to weak gravitational lensing convergence maps from the Subaru Hyper Suprime-Cam first-year data.
    \item[(2)] For the MFs, we show that the results approach the noiseless case for white noise levels of 0.1 or 1 $\mu$K-arcmin at 150 GHz. We also demonstrate that combining MFs measured with different smoothing scales does not improve the constraints.
    \item[(3)] For a realistic $f_{\rm sky}=0.4$, we find that the MFs and power spectrum respectively achieve an $\approx 0.04\%,$ and $0.06\%$ error bar on the best-constrained parameter combination $\Sigma_{8}=\sigma_{8}(\Omega_{c}/0.264)^{\alpha}$.  Here, $\alpha=0.11$ and $0.07$ for the MFs and power spectrum, respectively. 
\end{itemize}

We note that there are several simplifying assumptions in this analysis that would be important to study in follow-up work:
\begin{itemize}
    \item [(1)] \textit{Varying astrophysical parameters:} We fix all parameters associated with the ICM gas pressure profiles and focus only on the sensitivity to cosmological parameters. In practice, we would need to jointly fit the astrophysical parameters and cosmological parameters, since the ICM pressure profile is currently not well-constrained over the full mass and redshift range of interest here.  It would, therefore, be interesting and important to see how well these summary statistics can perform when astrophysical parameters are allowed to vary and if they can constrain some of the astrophysical parameters just as effectively. For example, Ref.~\cite{HillSherwin2013} used the different dependence of the variance and skewness on cosmological and astrophysical parameters to construct a statistic that cancels out the dependence on cosmology and can thus be used to constrain the astrophysical parameters. We can test this breaking of the degeneracies between ICM physics and cosmology by applying some of the statistics studied in this work. 
    
    \item [(2)] \textit{Effects of clustering:} We use simplified tSZ simulations that do not account for realistic clustering of halos in this work. Although clustering effects are expected to be small for tSZ statistics~\cite{KomatsuKitayama1999, HillPajer2013, Thiele2019}, it would be important to quantify this via a direct comparison with hydrodynamical or $N$-body simulations.

    \item [(3)] \textit{Realistic noise:} Another simplifying assumption in our noisy results is that we model both the foreground residuals in the tSZ maps after component separation and instrumental noise as GRFs using the estimated tSZ noise power spectra. In practice, there will be residuals in the maps that are non-Gaussian and correlated with tSZ signal (e.g., cosmic infrared background and point sources). In order to properly account for these effects, it will be important to forward model the astrophysical foregrounds and tSZ signal, processing the simulated maps through a component separation pipeline.
   
    \item [(4)] \textit{Gaussian likelihood assumption and constant covariance:} In this work, we assume a Gaussian likelihood for all observables. We have performed some checks on this assumption.  As an example, in the noiseless case, the distribution of the signal realizations at the fiducial cosmology in most MF bins are close to Gaussian-distributed. Similarly, Refs.~\cite{Gupta2018,Matilla2016,Lin2015} have shown that the Gaussian approximation is sufficient for many WL NG statistics. Nevertheless, it is important to directly quantify the effects of the Gaussian likelihood assumption on the tSZ constraints via, for example, simulation-based inference frameworks (e.g.,~\cite{DiazRivero2020, Novaes2024}). Since the tSZ simulations in this work are very inexpensive, such a comparison is not impeded by the availability of simulations and can thus be studied in great detail. Similarly, the effects of using a cosmology-dependent covariance should be quantified in future work. 
\end{itemize}

In summary, our results show that there is significant non-Gaussian information in the tSZ field, both in the noiseless and noisy cases. We find that the majority of this constraining power comes from halos that are not likely to be individually detected in upcoming tSZ cluster surveys, so the derived cosmological constraints are complementary to those found from tSZ cluster counts. This work is an important step toward characterizing the full information content of the tSZ field, with many interesting avenues for follow-up work.

\section{Data availability}

We make some data used in this analysis available at \url{https://columbialensing.github.io/} \cite{tSZ_maps}:
\begin{itemize}
    \item [(1)] The simulations at the fiducial cosmology and those with $\Omega_{c}, \sigma_{8}$ perturbed by $\pm1\%$ for Fisher forecasts. We share the full $f_{\rm sky}=0.1$ maps at the simulated 0.1 arcmin resolution (960 maps for the fiducial model and 142-144 for the four others).
    \item[(2)] Fiducial noiseless and noisy summary statistic data vectors ($816,696\times5\times2=8,166,960$ files).
\end{itemize}
We also make the main analysis pipeline available at \url{https://github.com/asabyr/tSZ_NG} \cite{tSZ_NG_code}. Due to the large storage requirements for the full simulation suite, all other simulations can be provided upon reasonable request to the corresponding author.

\section{Acknowledgements}
We are very grateful to Leander Thiele for assistance with \verb|hmpdf| and Boris Bolliet for help with \verb|class_sz|. We would also like to thank Shivam Pandey, Greg Bryan, Lam Hui, Sam Goldstein, Boris Bolliet, and Oliver Philcox for useful suggestions and discussions. We thank the anonymous referee for useful comments. AS and JCH acknowledge support from NASA grant 80NSSC23K0463 (ADAP).  ZH and JCH acknowledge support from NASA grant 80NSSC24K1093 (ATP).  JCH also acknowledges support from the Sloan Foundation and Simons Foundation.  We acknowledge the use of the NSF XSEDE facility for the simulations and data analysis in this study. We acknowledge computing resources from Columbia University's Shared Research Computing Facility project, which is supported by NIH Research Facility Improvement Grant 1G20RR030893-01, and associated funds from the New York State Empire State Development, Division of Science Technology and Innovation (NYSTAR) Contract C090171, both awarded April 15, 2010.
In addition to the software listed in the paper, we also make use of \verb|numpy|~\cite{numpy}, \verb|scipy|~\cite{scipy}, \verb|matplotlib|~\cite{matplotlib}, \verb|pixell|\footnote{\url{https://github.com/simonsobs/pixell}}, and \verb|colorgorical|~\cite{colorgorical} throughout this work.  This is not an official Simons Observatory Collaboration paper.

\appendix

\section{Numerical convergence of tSZ maps}\label{sec:convergence}

Our tSZ simulations draw the number of halos from a Poisson distribution by discretizing the halo mass function into bins of redshift and mass. Since the tSZ signal is dominated by the most massive halos, we must determine how many simulations are needed to capture the associated sample variance and if the mass function discretization settings are sufficient, before running the suite of different cosmologies. To do this, we simulate 5,184 maps using different choices for the number of bins in redshift ($N_{\rm z}$) and mass ($N_{\rm M}$) and look at the mean of each of the five summary statistics with respect to the number of simulations, $N_{\rm sims}$. 

\begin{figure*}[h]
    \centering
    \begin{minipage}{\textwidth}
        \centering
        \includegraphics[width=\textwidth]{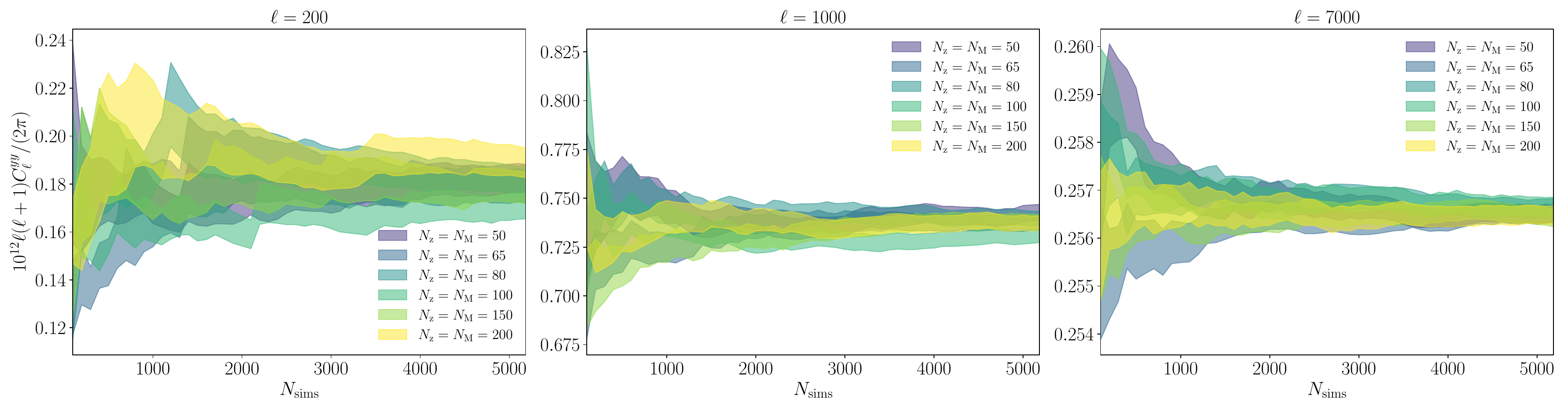}
    \end{minipage}
    \begin{minipage}{\textwidth}
        \centering
        \includegraphics[width=\textwidth]{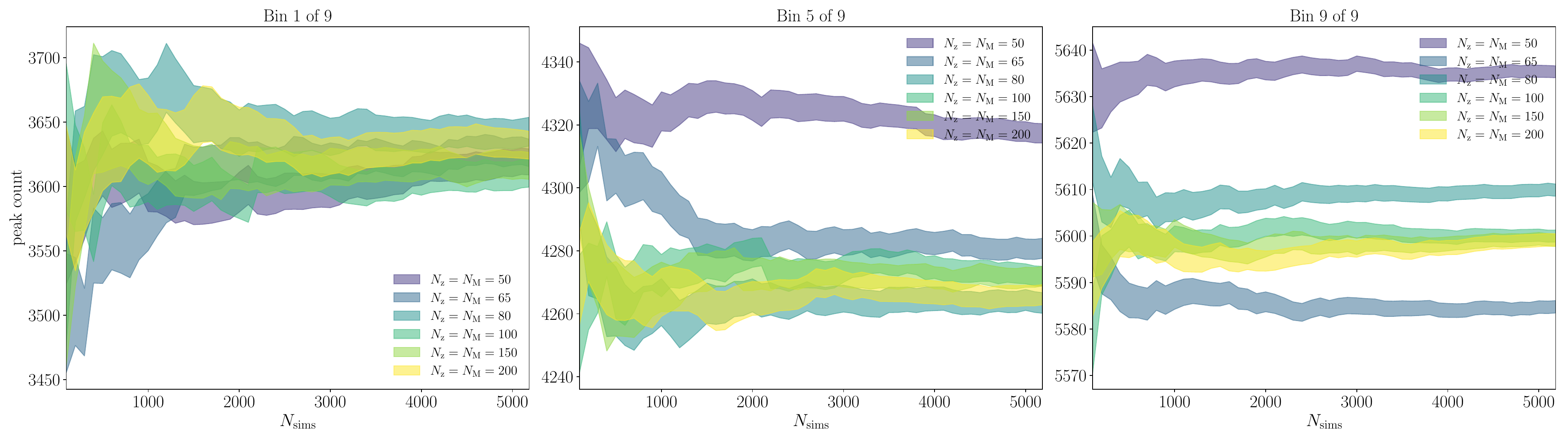}
        \label{fig:subfigB}
    \end{minipage}
    \begin{minipage}{\textwidth}
        \centering
        \includegraphics[width=\textwidth]{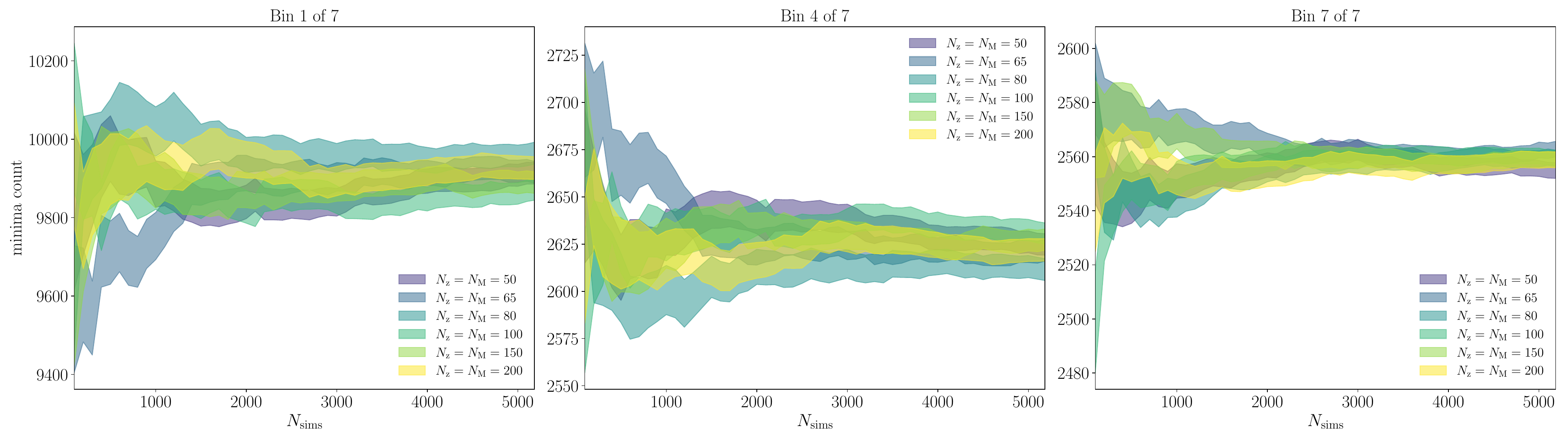}
    \end{minipage}
    \caption{Convergence of the mean of the different statistics with respect to the number of simulations and discretization of the halo mass function (i.e., $N_{\rm z}$ and $N_{\rm M}$ correspond to the number of bins in redshift and mass, respectively). \textit{Top:} Angular power spectrum at $\ell=200,1000,7000$ as a function of the numerical settings. \textit{Middle:} The convergence of some of the fiducial bins for peak counts. Note the discrepancy in peak counts for low $N_{\rm z}$ and $N_{\rm M}$. \textit{Bottom:} Convergence of the minima counts. \label{fig:stat_conv_1}}
\end{figure*}

The top panel in Fig.~\ref{fig:stat_conv_1} shows the angular power spectrum values for 3 of the 159 linearly-spaced $\ell$ bins centered between $\ell=\{50, 7950\}$, which are consistent between $N_{\rm z}=N_{\rm M}=150$ and $N_{\rm z}=N_{\rm M}=200$ at $N_{\rm sims}\sim5,000$ within the standard errors. 

In the second and third panels, we show the results for peak and minima counts. We determine the fiducial peak-count bin definitions by taking a realization of a single tSZ map and binning the peaks into 10 bins in such a way that every bin has an equal number of peaks. We merge the two highest-significance bins to ensure the peak counts are consistent within the standard errors for the three highest-precision numerical settings and end up with a total of 9 bins in our fiducial set-up. We check the three highest-precision numerical settings for this bin because of the more significant variation across different numerical settings compared to other observables. We show the first, fifth, and the last histogram bin in the figure. We see that the peak counts for the highest bin are significantly different when low values for $N_{\rm z}$ and $N_{\rm M}$ are used, but converge to the same mean for the higher values of $N_{\rm z}$ and $N_{\rm M}$. 

\begin{figure*}[h]
    \centering
    \includegraphics[width=0.7\textwidth]{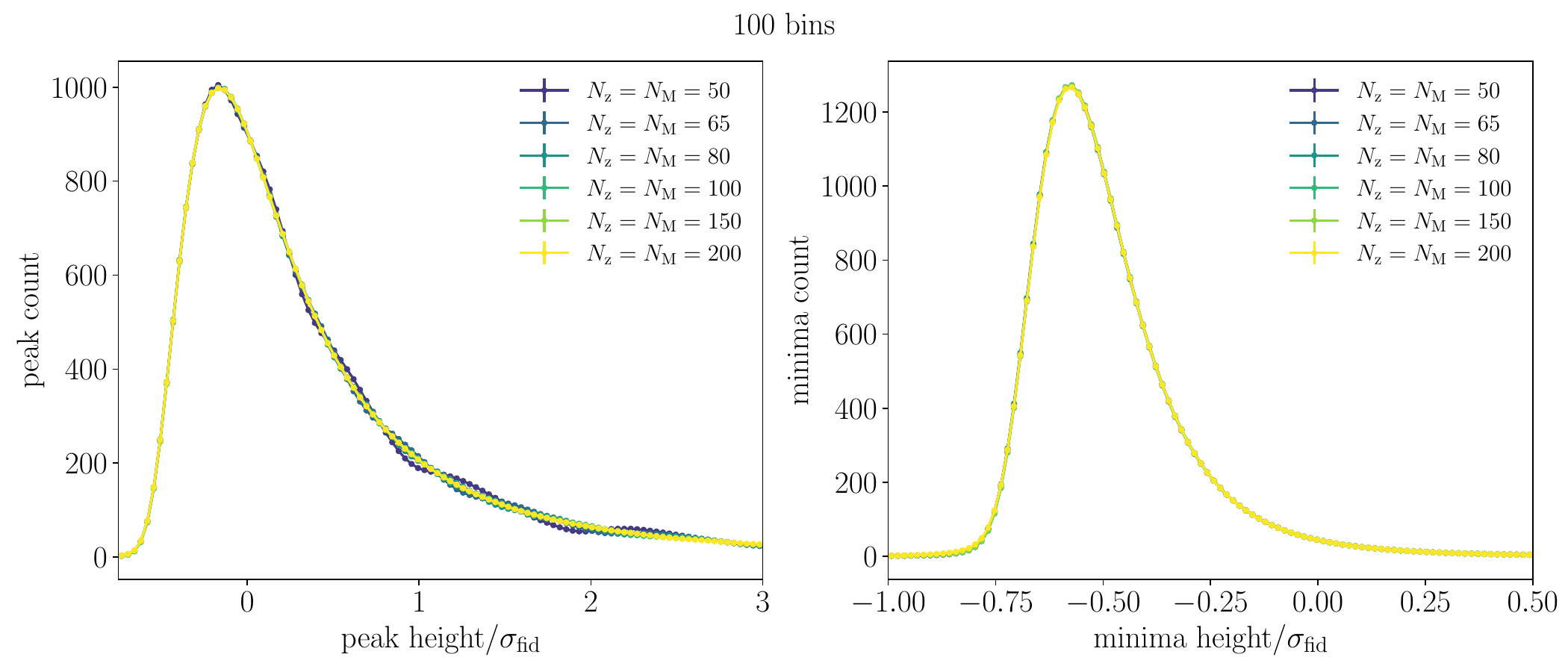}\caption{We show the finely binned peak (minima) histograms in the left (right) panel (100 linearly spaced bins). Note the numerical wiggles in the high-significance peak count bins when the halo mass function is not sufficiently discretized (i.e., when using low $N_{\rm z}$ and $N_{\rm M}$). \label{fig:100bins}}
\end{figure*}

Similarly, the bottom panel includes the results for minima counts. Since minima represent `valleys' in the tSZ field rather than possibly individual halos, we do not see such substantial effects. We follow a similar approach of determining 10 bins with equal number of minima and merging the bins to ensure convergence of the mean between $N_{\rm z}=N_{\rm M}=150$ and $N_{\rm z}=N_{\rm M}=200$. We adopt 7 bins in our fiducial set-up, three of which are shown in the figure. To further illustrate the effects of numerical discretization, in Fig.~\ref{fig:100bins} we plot the peak and minima counts for 100 linearly-spaced bins between $\{-0.75,3\}\sigma_{\rm fid}$ and $\{-1,0.5\}\sigma_{\rm fid}$, respectively. The peak histogram clearly shows wiggles in the high-significance tail due to insufficient discretization at low values of $N_{\rm z}$ and $N_{\rm M}$. 

For the MFs, we show one bin for each of $V_{0}$, $V_{1}$, and $V_{2}$. We check that all the bins for each $V_{0}$, $V_{1}$, and $V_{2}$ are consistent between $N_{\rm z}=N_{\rm M}=150$ and $N_{\rm z}=N_{\rm M}=200$ within one or two standard errors. We also show the convergence for some of the moments and check that they are converged to the same mean. Based on these results, we adopt the numerical settings $N_{\rm z}=N_{\rm M}=150$ and conclude that $\sim 5000$ maps suffices for each cosmology.


\begin{figure*}[h]
\begin{minipage}{\textwidth}
        \centering
        \includegraphics[width=\textwidth]{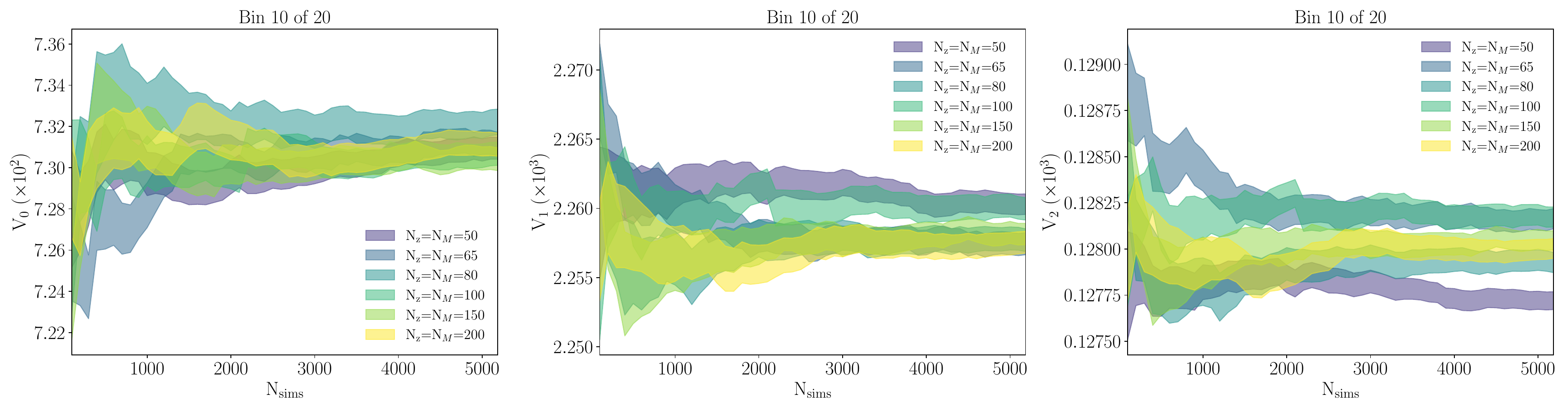}
    \end{minipage}
    \begin{minipage}{\textwidth}
        \centering
    \includegraphics[width=\textwidth]{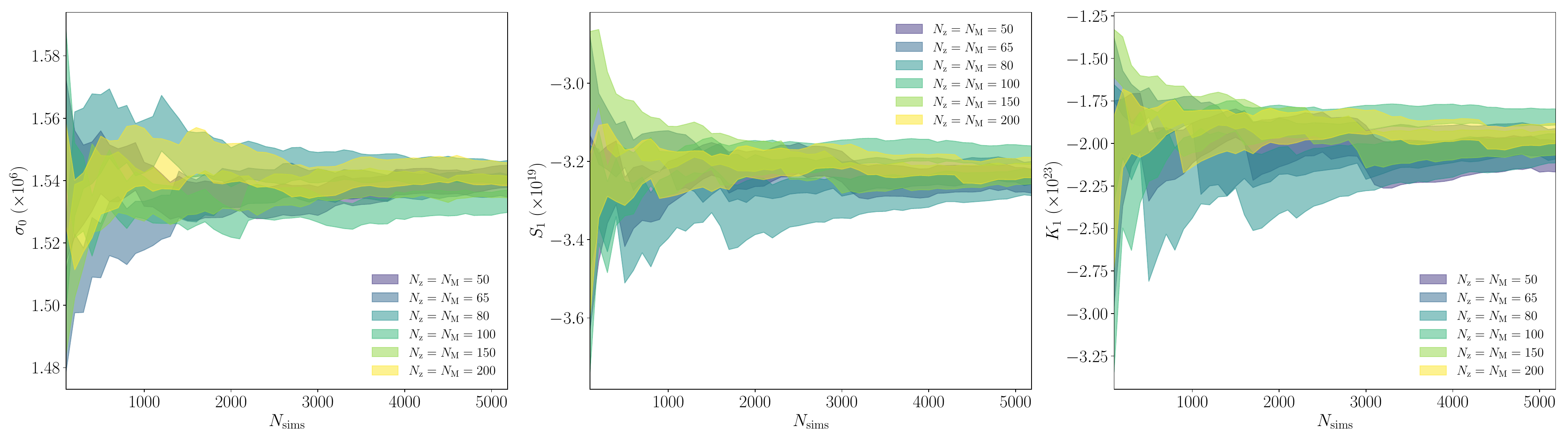}
    \end{minipage}
    \label{fig:stat_conv_2}
    \caption{Convergence of the mean value of MFs and moments with respect to the number of simulations and the discretization of the halo mass function (i.e., $N_{\rm z}$ and $N_{\rm M}$ correspond to the number of bins in redshift and mass, respectively).}
\end{figure*}

\begin{figure*}[h]
    \centering
    \includegraphics[width=\textwidth]{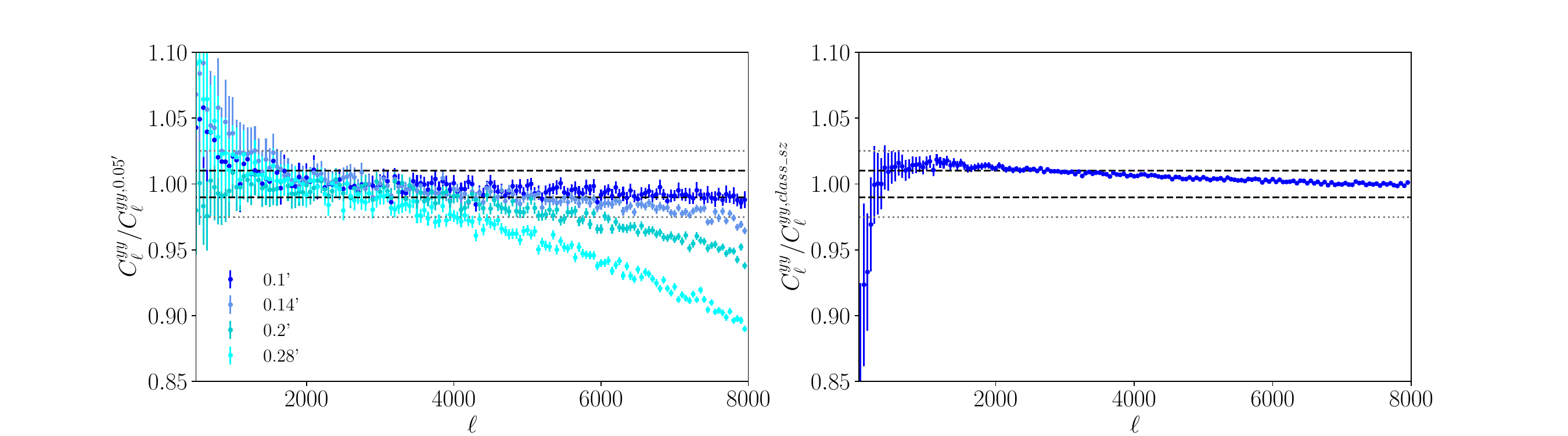}[h]
    \cprotect\caption{Validation of the simplified tSZ simulations via power spectrum comparisons. The black (grey) dashed lines indicate $1\%$ ($2.5\%$) difference levels. \textit{Left:} The ratio between the average of 300 power spectra computed from maps with different resolutions to the power spectrum from the maps at 0.05 arcmin resolution, when smoothed using a Gaussian kernel with $\thetagauss = 1.4$ arcmin. \textit{Right:} Comparison of the  estimated power spectra for our fiducial cosmology to that computed from the halo model code \verb|class_sz|. The power spectra are estimated using 5,112 realizations of $10.5\times10.5$ deg$^{2}$ maps.}
    \label{fig:cl_validation}
\end{figure*}

Additionally, we compare the power spectra computed from the maps generated at different resolutions to determine the lowest map resolution that we can use without losing the accuracy of the signal, given the $\thetagauss=1.4$ arcmin beam smoothing, which we apply in all the forecasts. The left panel in Fig.~\ref{fig:cl_validation} shows the ratios for power spectra computed from several sets of maps at different resolutions compared to the power spectrum at 0.05 arcmin resolution. For this comparison, we use 300 realizations of $\sim10\times10$ deg$^{2}$ maps, remove the square pixel window function, and multiply each power spectrum by the mean of the apodization mask, or $\sum_{n=1}^{n=N} M(n)/N$, where $M$ is the apodization mask value at each pixel $n$ and $N$ is the total number of pixels in the given map (i.e., approximating that there is an equal suppression of power across all scales due to the apodization mask). We have checked that this simple multiplicative factor is sufficient in this case by comparing to the power spectra estimated with the full  MASTER pseudo-$C_\ell$ approach for masked fields implemented in \verb|NaMaster|\cite{namaster}\footnote{\url{https://github.com/LSSTDESC/NaMaster}}. Based on the results in Fig.~\ref{fig:cl_validation}, we conclude that 0.1 arcmin resolution is sufficient. 

To check the accuracy of the simulated tSZ signal, we compare the power spectrum from the maps against the halo-model code \verb|class_sz|\cite{class_sz_mm_paper,Bolliet2018,Bolliet2020,Bolliet2023,KomatsuSeljak2002}. In this comparison, we show the results averaged over 5,112 map realizations. We divide out the pixel window function, account for the apodization mask, and multiply the theoretical halo model prediction by the same Gaussian beam. The power spectra are in good agreement (close to percent level at most $\ell$) as shown in the right panel of Fig.~\ref{fig:cl_validation}. 

\section{White noise}
\begin{figure*}[h]
    \centering
    \includegraphics[width=\textwidth]{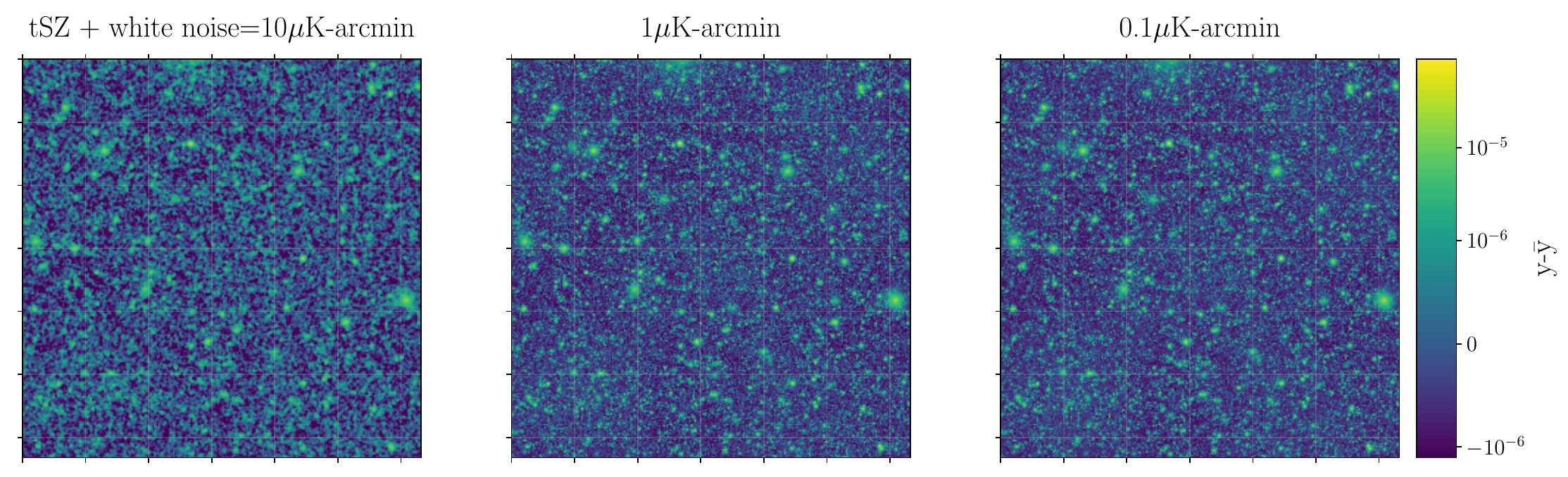}
    \caption{Simulated tSZ map combined with a Gaussian random field generated from a flat (white) power spectrum with noise levels corresponding to 10, 1, and 0.1 $\mu$K-arcmin at $\nu=150$ GHz. In the cases of 1 or 0.1 $\mu$K-arcmin white noise, the field closely resembles the noiseless map.}
    \label{fig:white_noise}
\end{figure*}

In Fig. \ref{fig:white_noise}, we show tSZ maps combined with different levels of white noise (see \S~\ref{sec:sims} for a description).  

\section{Additional noiseless constraints tests}\label{app:interp_bin}

In addition to testing the stability of the noiseless constraints to the number of simulations used for the covariance or mean summary statistic estimations described in \S~\ref{sec:noiseless_convergence}, we test the constraints against interpolation errors and fiducial binning choices. 

We check that the errors from interpolating the summary statistics at the cosmologies where we do not have simulations are negligible. We do this by testing the accuracy of the interpolator by excluding the fiducial cosmology in its construction (i.e., using 153 cosmologies) and checking the difference in the predicted observable values versus those measured directly from the fiducial cosmology. We find
that the difference between interpolated and true values are all $< 0.03\sigma$ with a median of $0.01\sigma$, where $\sigma$ refers to the rms in each observable bin. We also check that the contours remain the same if we only use a subset of cosmologies. For example, we compute the contours using only 102 cosmologies from the initial LH suite (see \S\ref{sec:sims}) and confirm that the contours remain the same (i.e., $\sim1\%$ difference in contour area in the case of using all summary statistics combined). With these tests, we conclude that any errors from interpolation are minor.

We check how stable our results are to binning choices for the MFs, peaks, and minima. For peaks and minima, we do so by computing the constraints using different bin edges. For this exercise, we keep the outermost bins the same (e.g., $[-100, -0.3, ... 1, 1000]\sigma_{\rm fid}$), while combining or splitting the rest of the bins. We find that for peaks, the results remain stable and almost the same regardless of the number of bins ($<1\%$ difference using 30 or 16 versus 9 bins, and $\sim7\%$ change when using 4 bins). The contour area is only significantly impacted when we use a single bin for the peak counts. For the minima counts, using 4, 12, or 22 bins similarly leads to small differences in the area: $\sim18\%, 3\%$, and $4\%$ increase, respectively. The contour sizes are most stable for 12 and 22 bins. For the MFs, we find that the areas are stable when increasing the number of bins to 50, 80, or 100. From these tests, we conclude that the constraints presented in \S~\ref{sec:noiseless} are stable to these choices.

\section{Different maximum mass cutoffs: covariance and correlation matrices}
\label{app:cutoffs}

\begin{figure*}[h]
    \centering
    \includegraphics[width=\linewidth]{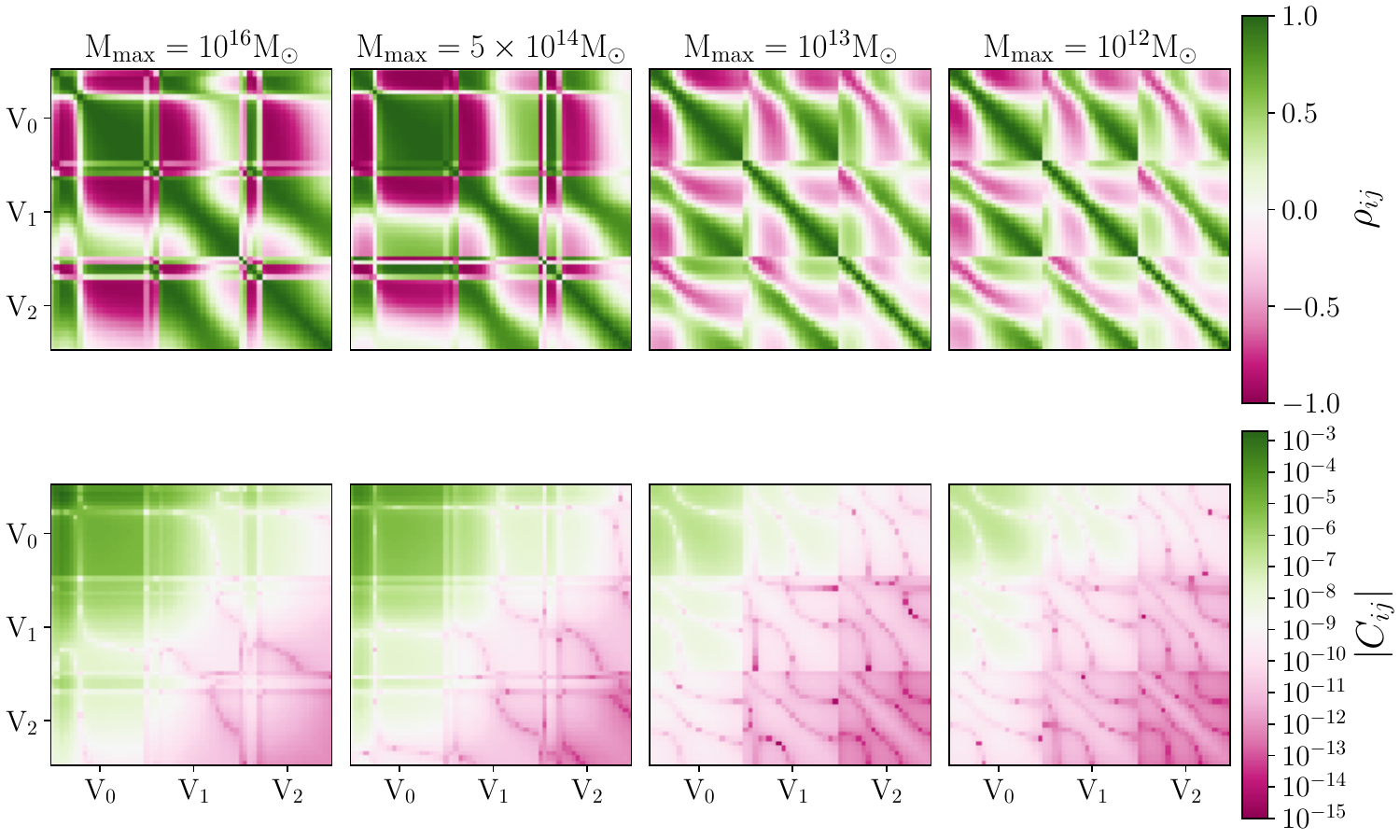}
    \caption{\emph{Top}: Correlation matrices computed for the fiducial cosmology (34,560 realizations) using simulations with different maximum halo mass cutoffs. Excluding massive halos reduces the correlation between the MF bins. \emph{Bottom}: Absolute covariance matrices for the different mass cutoffs. The covariance is similarly reduced as the maximum halo mass used in the simulation is lowered.}
    \label{fig:mass_corr}
\end{figure*}

In Fig.~\ref{fig:mass_corr}, we show the correlation and absolute values of the covariance matrices for the MFs computed from simulations with different maximum halo mass cutoffs, as studied in \S~\ref{sec:noiseless} and in Fig.~\ref{fig:mass_corr}. Both correlation across the MF bins and the total covariance are lowered when the most massive halos are excluded from the maps.

\section{Additional moments}\label{app:kurtosis}
The full set of moments required to write down the perturbative expansion of the MFs up to second order in variance consists of nine moments. In the main text, we have restricted our results to a total of four moments by including only one cubic and one quartic moment due to possible lack of convergence of the rest of the cubic and quartic moments ($S_{0}$, $S_{2}$, $K_{0}$, $K_{2}$, $K_{3}$):

\begin{equation}
S_{0}=\langle y^{3}\rangle, \quad S_{2}=\langle|\nabla y|^{2}\nabla^{2}y\rangle
\end{equation}
\begin{equation}
K_{0}=\langle y^{4} \rangle, \quad K_{2}=\langle y \nabla^{2} y |\nabla y|^{2} \rangle, \quad K_{3}=\langle|\nabla y|^{4}\rangle 
\end{equation}

\begin{figure*}[h]
    \centering
    \includegraphics[width=0.7\textwidth]{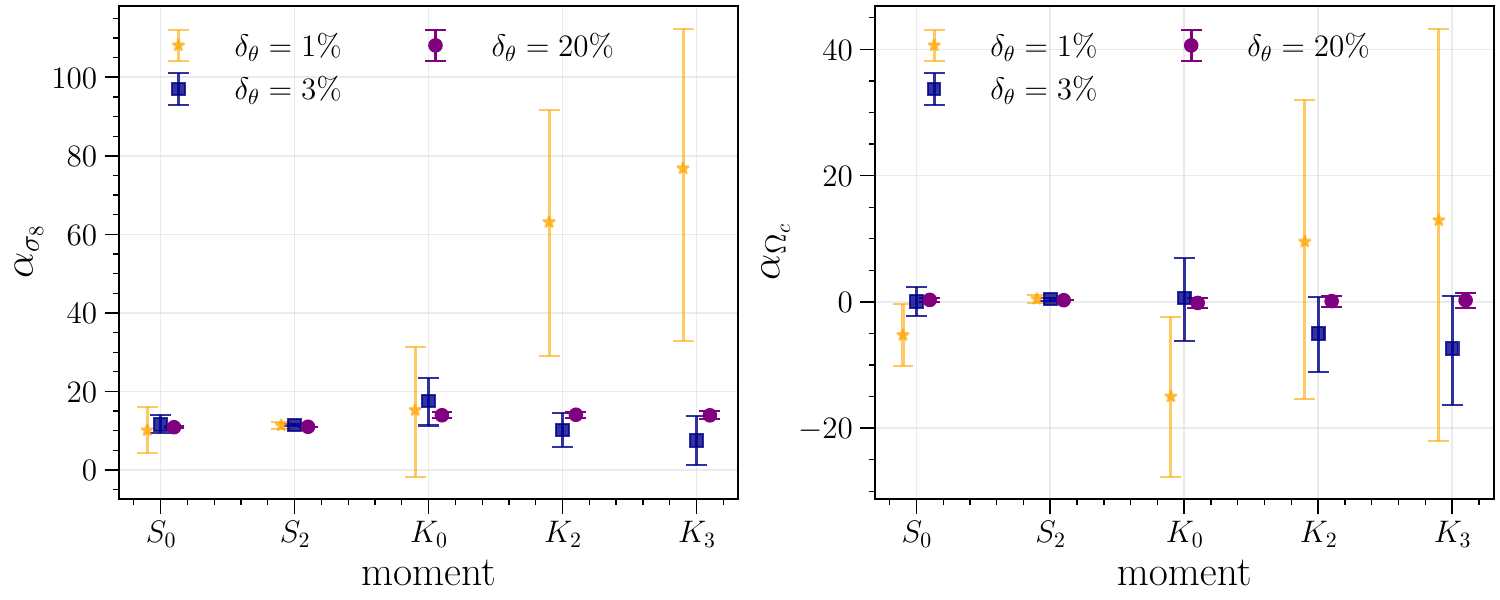}
    \caption{Same as Fig.~\ref{fig:scalings} for the additional two cubic and three quartic moments.}
    \label{fig:rest_scalings}
\end{figure*}

\begin{figure*}[h]
    \centering
    \includegraphics[width=\textwidth]{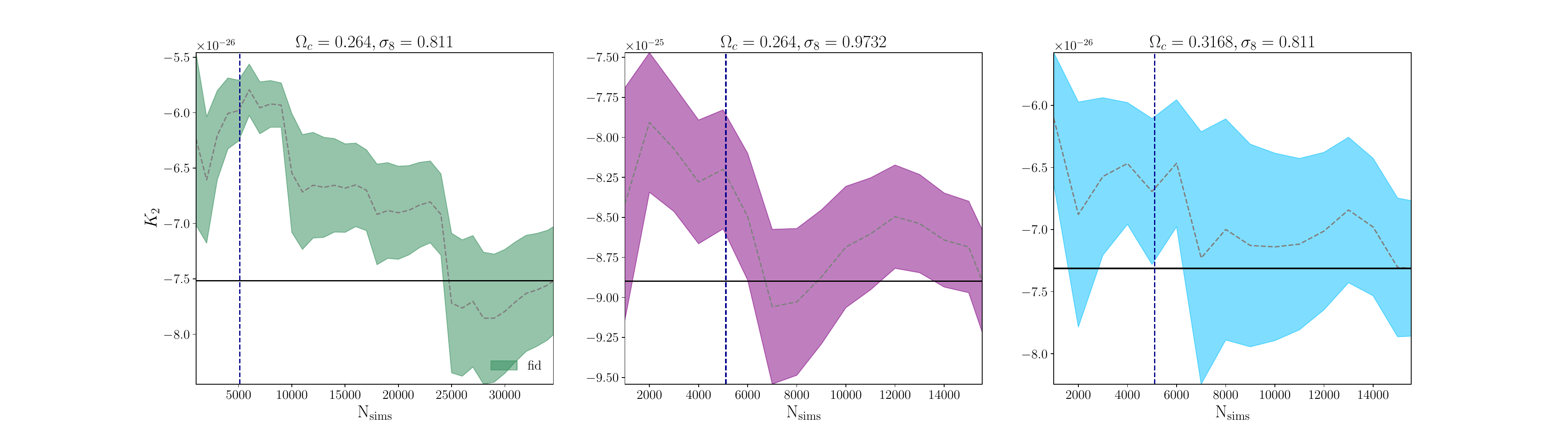}
    \caption{Convergence of the $K_{2}$ moment as a function of the number of simulations for three different cosmologies: fiducial and two example cosmologies with $\sigma_{8}$ and $\Omega_{c}$ parameters increased by $1.2\times$. We show the mean at each number of simulations (dashed grey), and the mean at the maximum number of simulations, $N_{\rm fid}=34,560$ and $N_{\rm sim}=15,552$ (black solid). The vertical line corresponds to 5,112 simulations, which is the number of simulations used for the mean value of the summary statistics at all cosmologies except the fiducial.}
    \label{fig:conv_K2}
\end{figure*}

In Fig. \ref{fig:conv_K2}, we show an example convergence plot for one of these moments, $K_{2}$, for three cosmologies: fiducial and those with $\sigma_{8}$ and $\Omega_{c}$ increased by a factor of $1.2$. In these plots, we show the convergence of the mean value using 34,560 simulations for the fiducial and $15,552$ for the example cosmologies. The plot shows that $K_{2}$ value is likely unconverged using 5,112 sims at each of the cosmologies in the suite. For completeness, we include a table with the values for the rest of these moments at the three cosmologies and a plot of their scalings with cosmological parameters in Table \ref{tab:rest_example_moments} and Fig.~\ref{fig:rest_scalings}, similar to what we have shown in \S\ref{sec:noiseless}. 

In Table~\ref{tab:rest_constraints}, we summarize the results we obtain when we include these additional moments both in the baseline noiseless and noisy cases. We also show the constraints for each moment individually, except $\sigma_{0}$, which is not constraining on its own. In the noiseless case, we find that by including all nine moments, we still obtain constraints weaker than those from the power spectrum, with $A_{68}^{\rm ps}/A_{68}=0.79$. But the constraints are $\approx1.1\times$ tighter than when using only the four moments considered in the main text. In the noisy case, we find that the constraints from all moments are $\approx 1.1\times$ better than the power spectrum alone and $\approx1.1\times$ better than when using just the four moments. Looking at the moments individually, we see that in the noiseless case, the constraints are improved the most by inclusion of each new higher-order moment of the derivatives (e.g., when $S_{1}$ or $K_{1}$ are added), while adding additional higher-order moments with or without the derivatives has a less substantial effect. In the noisy case, we see similar behavior, but adding additional higher-order moments of the derivatives has larger effects (e.g., adding $S_{2}$ or $K_{2}$). In the future, it may be advantageous to explore various ways of Gaussianizing the tSZ field by applying a non-linear transformation to the maps before measuring these moments to more easily reach convergence of their mean values and take advantage of the additional constraining power from these moments (e.g., \cite{Neyrinck2009, Carron2013trans}). We leave this exploration to follow-up work.

\begin{table*}[h]
    \centering
    \begin{tabular}{|c|c|c|c|c|c|}
    \colrule
     & $S_{0}\times10^{-17}$ & $S_{2}\times10^{-22}$ & $K_{0}\times10^{-21}$ & $K_{2}\times10^{-26}$ & $K_{3}\times10^{-25}$\\
    \colrule
    fiducial& 5.74&$-8.19$& 4.18& $-7.52$&1.10\\
    $1.2\times\sigma_{8}$ & 41.7&
 $-60.9$ & 53.1& $-97.5$&13.9\\
  $\alpha_{\sigma_{8}}$ &10.9&11.0&13.9&14.1&13.9\\
    $1.2\times\Omega_{c}$ & 6.07&$-8.59$& 4.06& $-7.69$&1.15\\
    $\alpha_{\Omega_{c}}$&0.3&0.3&$-0.2$&0.1&0.2\\
    \colrule
    \end{tabular}
    \caption{Same as Table~\ref{tab:example_moments}, but for the five additional moments not considered in the main text.}
    \label{tab:rest_example_moments}
\end{table*}

\begin{table*}[h]
    \centering
    \begin{tabular}{|l|c|c|c|}
    \colrule
         Observable & $A_{68}$ $\times10^{4}$ & $A_{68}^{\rm ps}$/$A_{68}$ & $\hat{\sigma_{8}}$\\
        \colrule
        Noiseless & & & \\
        $\sigma_{0}$, $\sigma_{1}$ &16.9&0.46&$0.81_{-0.01}^{+0.011}$\\
        $\sigma_{0}$, $\sigma_{1}$, $S_{0}$ & 16.9&0.46&$0.81_{-0.01}^{+0.011}$\\
        $\sigma_{0}$, $\sigma_{1}$, $S_{0}$, $S_{1}$&12.6&0.62&$0.81_{-0.01}^{+0.011}$\\
        $\sigma_{0}$, $\sigma_{1}$, $S_{0}$, $S_{1}$, $S_{2}$&12.5&0.63&$0.81\pm0.01$\\
        $\sigma_{0}$, $\sigma_{1}$, $S_{0}$, $S_{1}$, $S_{2}$, $K_{0}$&12.1&0.65&$0.81\pm0.01$\\
        $\sigma_{0}$, $\sigma_{1}$, $S_{0}$, $S_{1}$, $S_{2}$, $K_{0}$, $K_{1}$ &9.96&0.79&$0.81_{-0.01}^{+0.011}$\\
        $\sigma_{0}$, $\sigma_{1}$, $S_{0}$, $S_{1}$, $S_{2}$, $K_{0}$, $K_{1}$, $K_{2}$&9.94&0.79&$0.81_{-0.01}^{+0.011}$\\
        $\sigma_{0}$, $\sigma_{1}$, $S_{0}$, $S_{1}$, $S_{2}$, $K_{0}$, $K_{1}$, $K_{2}$, $K_{3}$ &9.86&0.79&$0.81\pm0.01$\\
         \colrule
         Noisy & & & \\
        $\sigma_{0}$, $\sigma_{1}$ &18.3&0.67&$0.809_{-0.01}^{+0.012}$\\
        $\sigma_{0}$, $\sigma_{1}$, $S_{0}$ & 18.3&0.67&$0.809_{-0.01}^{+0.012}$\\
        $\sigma_{0}$, $\sigma_{1}$, $S_{0}$, $S_{1}$&14.1&0.87&$0.810_{-0.01}^{+0.011}$\\
        $\sigma_{0}$, $\sigma_{1}$, $S_{0}$, $S_{1}$, $S_{2}$&13.2&0.93&$0.810\pm0.011$\\
        $\sigma_{0}$, $\sigma_{1}$, $S_{0}$, $S_{1}$, $S_{2}$, $K_{0}$&13.2&0.93&$0.810_{-0.01}^{+0.011}$\\
        $\sigma_{0}$, $\sigma_{1}$, $S_{0}$, $S_{1}$, $S_{2}$, $K_{0}$, $K_{1}$ &11.9&1.04&$0.810_{-0.01}^{+0.011}$\\
        $\sigma_{0}$, $\sigma_{1}$, $S_{0}$, $S_{1}$, $S_{2}$, $K_{0}$, $K_{1}$, $K_{2}$&11.2&1.1&$0.810\pm0.011$\\
        $\sigma_{0}$, $\sigma_{1}$, $S_{0}$, $S_{1}$, $S_{2}$, $K_{0}$, $K_{1}$, $K_{2}$, $K_{3}$ &11.1&1.1&$0.810\pm0.011$\\
         \colrule
    \end{tabular}
    \caption{Summary of the constraints obtained using nine moments in both the noiseless and noisy cases.}
    \label{tab:rest_constraints}
\end{table*}

\bibliography{main}
\end{document}